\documentclass[twocolumn, aps, superscriptaddress, prb, showpacs,floatfix,longbibliography]{revtex4-1}
\usepackage{dcolumn}
\usepackage{graphicx}
\usepackage{amsmath,amssymb}
\usepackage{bm,xcolor,url}
\usepackage{pict2e,hyperref}

\usepackage{here}

\newcommand{\ket}[1]{|{#1}\rangle}
\newcommand{\bra}[1]{\langle{#1}|}

\begin{document}
\title {Bulk-entanglement spectrum correspondence in $PT$- and $PC$-symmetric topological insulators and superconductors}
\author {Ryo Takahashi}
\affiliation{
Advanced Institute for Materials Research, Tohoku University, Sendai 980-8577, Japan\\
}
\author {Tomoki Ozawa}
\affiliation{
Advanced Institute for Materials Research, Tohoku University, Sendai 980-8577, Japan\\
}

\date{\today}
\begin{abstract}
In this study, we discuss a new type of bulk-boundary correspondence which holds for topological insulators and superconductors when the parity-time ($PT$) and/or parity-particle-hole ($PC$) symmetry are present. In these systems, even when the bulk topology is nontrivial, the edge spectrum is generally gapped, and thus the conventional bulk-boundary correspondence does not hold. We find that, instead of the edge spectrum, the single-particle entanglement spectrum becomes gapless when the bulk topology is nontrivial: i.e., the {\it bulk-entanglement spectrum correspondence } holds in $PT$- and/or $PC$-symmetric topological insulators and superconductors. After showing the correspondence using $K$-theoretic approach, we provide concrete models for each symmetry class up to three dimensions where nontrivial topology due to $PT$ and/or $PC$ is expected. An implication of our results is that, when the bulk topology under $PT$ and/or $PC$ symmetry is nontrivial, the non-interacting many-body entanglement spectrum is multiply degenerate in one dimension and is gapless in two or higher dimensions. 
\end{abstract}

\maketitle
\section{Introduction}
Recently emerging platforms of two dimensional materials such as twisted bilayer graphene and other van der Waals layered materials~\cite{Novoselov:2016,Liu:2016,Andrei:2020} have led to interests in understanding properties of materials involving spatial inversion, i.e., parity symmetry. Indeed, new classes of two-dimensional topological insulators such as the Euler insulators and Stiefel-Whitney insulators are proposed~\cite{PhysRevB.92.081201,PhysRevLett.118.056401,PhysRevLett.121.106403,PhysRevX.9.021013,PhysRevLett.125.053601,PhysRevB.102.115135,PhysRevB.103.205303,zhao2022quantum}, where the topology is protected by the $PT$ symmetry and the topological invariants are distinct from more conventional Chern numbers and $\mathbb{Z}_2$ topological invariants~\cite{RevModPhys.82.3045,RevModPhys.83.1057}. Here, $P$ is the parity symmetry and $T$ is the time-reversal symmetry; the topology of $PT$-symmetric insulators are protected as long as the product of these two symmetries are kept. Further extension and generalization of such topological insulators and superconductors have been found. For example, $PC$ symmetry, where $C$ is the particle-hole symmetry, can also lead to new classes of topological phases~\cite{PhysRevB.90.024516,PhysRevLett.116.156402,PhysRevB.96.155105}. In the context of engineered quantum materials, such $PT$- and $PC$-symmetric topological insulators have also been studied experimentally, revealing various unique topological features~\cite{PhysRevLett.125.053601,jiang2021experimental,zhao2022quantum,peng2022phonons,xue2023stiefel}. 

One distinct feature of $PT$- and $PC$-symmetric topological insulators and superconductors is the lack of the ordinary bulk-boundary correspondence: i.e., the absence of edge physics reflecting the nontrivial bulk topology. Even when the bulk topology is nontrivial, the edge spectrum is generally gapped in such systems. This violation of the bulk-boundary correspondence is because the $PT$ and $PC$ symmetries are generally broken at the edge of the system due to the spatially non-local nature of the parity operation.  The lack of manifest edge physics in $PT$- or $PC$-symmetric topological insulators and superconductors puts severe constraints on experimental access to the topological physics of these materials. Indeed, so far the physical manifestation of such $PT$- or $PC$-symmetric topological insulators and superconductors are restricted to direct detection of the bulk properties of the material~\cite{PhysRevLett.125.053601,jiang2021experimental,zhao2022quantum,peng2022phonons}. Given the important role the edge physics played in the case of ordinary topological insulators and superconductors, it is desirable to find topological features of $PT$- or $PC$-symmetric topological insulators and superconductors related to edge properties.

In previous works, parity-symmetric topological insulators were found to have no direct bulk-edge correspondence but instead show distinct features in the so-called entanglement spectrum. entanglement spectrum is a property of the reduced density matrix of the system when the system is cut in two (or more) and integrating out a part of the system~\cite{PhysRevLett.101.010504,PhysRevB.81.064439,zeng2019quantum}. The full many-body entanglement spectrum of non-interacting fermions can be computed from the so-called single-particle entanglement spectrum, which is essentially the spectrum of a correlation matrix~\cite{Ingo_Peschel_2003,PhysRevA.67.052311,PhysRevLett.104.130502,peschel2012entanglement}. It was found that when the bulk topology is nontrivial in the presence of the parity symmetry, the single-particle entanglement spectrum is gapless around zero energy, and the corresponding eigenvectors of these gapless modes are localized around the cut of the system introduced to define the entanglement~\cite{PhysRevB.82.241102,PhysRevB.83.245132,PhysRevB.84.195103,PhysRevB.85.165120}.

We find that a similar correspondence between the bulk topology and the single-particle entanglement spectrum holds for a much more general setup of all symmetry classes of $PT$- and $PC$-symmetric topological insulators and superconductors in all dimensions. The $PT$- and $PC$-symmetric topological insulators and superconductors are classified into ten distinct classes called AZ$+I$ classification~\cite{PhysRevB.96.155105} according to the presence and absence of $PT$ and $PC$ symmetries. In our previous paper, we have discussed the bulk-entanglement spectrum correspondence in two dimensions when only $PT$ with the property $(PT)^2 = +1$ is present, which is the class AI$^\prime$ in the AZ$+I$ classification~\cite{PhysRevB.108.075129}. The goal of this paper is to show that this correspondence can be extended to all classes and dimensions.

Here is the outline of this paper. In Sec.~\ref{sec:classification}, we review and summarize the classification of the bulk topology and gapless modes of both AZ and AZ$+I$ classification from the perspective of $K$-theory. Using the results of the classification, in Sec.~\ref{sec:bbc}, we prove the bulk-entanglement spectrum correspondence for all nontrivial classes of the AZ$+I$ classification. In Sec.~\ref{sec:models}, we construct models for all nontrivial $\mathbb{Z}_2$ classes of the AZ$+I$ classification up to three spatial dimensions, and explicitly confirm the bulk-entanglement spectrum correspondence. We note that for symmetry classes described by $\mathbb{Z}$ or $2\mathbb{Z}$ topological invariants, the bulk-entanglement spectrum correspondence reduces to the ordinary bulk-edge correspondence. In Sec.~\ref{sec:fragile}, we also extend our analysis of the bulk-entanglement spectrum correspondence for fragile insulators, which are topological insulators not robust against adding and mixing trivial bands below the Fermi level.
In Sec.~\ref{sec:many-body}, we relate the single-particle entanglement spectrum to the many-body entanglement spectrum to find implications of the bulk-entanglement spectrum correspondence on the multiplicity and the gapless nature of the many-body entanglement spectrum. We finally give conclusion and future prospects in Sec.~\ref{sec:conclusion}. Details of some calculations are given in the Appendix of the paper.

\section{Classification of topological phases and stable gapless modes}
\label{sec:classification}
In this section, we briefly review and summarize the classification of topological phases and stable gapless modes under the symmetry classes of the AZ and AZ$+I$ classification. Although our goal is to understand the entanglement spectrum under the symmetry classes of AZ$+I$ classification, here we discuss both AZ and AZ$+I$ classifications because many of the arguments go in parallel; in this way, similarities and differences between AZ and AZ$+I$ become more evident. 

The AZ classification is a symmetry classification based on the presence or absence of three internal symmetries: time-reversal ($T$), particle-hole ($C$), and chiral (or sublattice) symmetry ($\Gamma$). 
Here, $T$ and $C$ are antiunitary symmetries, whereas $\Gamma$ is a unitary symmetry. Presence of each symmetry implies the following conditions on the Bloch Hamiltonian $H(\mathbf{k})$:
\begin{align}
TH(\mathbf{k})T^{-1} &= H(-\mathbf{k}), &
CH(\mathbf{k})C^{-1} &= -H(-\mathbf{k}),
\\
\Gamma H(\mathbf{k}) \Gamma^{-1}&=- H(\mathbf{k}). 
\end{align}
Depending on the presence or absence of these three symmetries, there are ten different classes of Bloch Hamiltonians as shown in Table~\ref{tab:classes}.
Symmetry classes without $T$ and $C$ are called complex classes, which are classes A and AIII in the table.
On the other hand, classes which possess at least one of $T$ or $C$ are called real classes. There are eight real classes, and we label them with an integer $s$ (mod $8$) as in the table. 

\begin{table}[tb]
\begin{center}
\caption{Definition of (left) AZ and (right) AZ$+I$ classes. In columns under $T$, $C$, $PT$, and $PC$, $+$ denotes that the square of the corresponding symmetry is $+1$, whereas $-$ means that the square is $-1$, and 0 means the absence of the corresponding symmetry. In columns under $\Gamma$, 1 denotes the existence and 0 denotes the absence of the chiral symmetry $\Gamma$.}
\begin{tabular}{c|c|cccc}
     $s$ &   \phantom{+}AZ\phantom{+} & \phantom{I}$T$\phantom{I} & \phantom{I}$C$\phantom{I} & \phantom{I}$\Gamma$\phantom{I} \\  \hline \hline
     0 &   A & 0 & 0 & 0   \\
     1 &  AIII & 0 & 0 & 1  \\ \hline
     0 &   AI & $+$ & 0 & 0 \\
     1 &  BDI & $+$ & $+$ & 1 \\
     2 &    D & 0 & $+$ & 0 \\
     3 & DIII & $-$ & $+$ & 1 \\
     4 &  AII & $-$ & 0 & 0 \\
     5 &  CII & $-$ & $-$ & 1 \\
     6 &    C & 0 & $-$ & 0 \\
     7 &   CI & $+$ & $-$ & 1 \\
\end{tabular}
\quad \quad
\begin{tabular}{c|c|cccc}
     $s$ &   \phantom{,}AZ$+I$\phantom{,} & $PT$ & $PC$ & \phantom{I}$\Gamma$\phantom{I} \\  \hline \hline
     0 &   A & 0 & 0 & 0   \\
     1 &  AIII & 0 & 0 & 1  \\ \hline
     0 &   AI$^\prime$ & $+$ & 0 & 0 \\
     1 &  BDI$^\prime$ & $+$ & $+$ & 1 \\
     2 &    D$^\prime$ & 0 & $+$ & 0 \\
     3 & DIII$^\prime$ & $-$ & $+$ & 1 \\
     4 &  AII$^\prime$ & $-$ & 0 & 0 \\
     5 &  CII$^\prime$ & $-$ & $-$ & 1 \\
     6 &    C$^\prime$ & 0 & $-$ & 0 \\
     7 &   CI$^\prime$ & $+$ & $-$ & 1 \\
\end{tabular}
\label{tab:classes}
\end{center}
\end{table}

The AZ$+I$ classification is a symmetry classification in which the antiunitary symmetries $T$ and $C$ of the AZ classification are replaced by $PT$ and $PC$, which are composite operations with the spatial inversion operation $P$. The definition of the chiral symmetry $\Gamma$ remains the same. The presence of each symmetry implies the following conditions on the Bloch Hamiltonian $H(\mathbf{k})$:
\begin{align}
(PT)H(\mathbf{k})(PT)^{-1} &= H(\mathbf{k}), &
(PC)H(\mathbf{k})(PC)^{-1} &=-H(\mathbf{k}). 
\end{align}
We note that, unlike $T$ and $C$ operators alone, the combinations $PT$ and $PC$ do not change the sign of $\mathbf{k}$. 
Also in the AZ$+I$ classification, depending on the presence or absence of $PT$, $PC$, and $\Gamma$, Bloch Hamiltonians are classified into ten classes as shown in Table~\ref{tab:classes}.
We note that classes A and AIII are the same between AZ and AZ$+I$ classifications, because the chiral symmetry $\Gamma$ is common for both AZ and AZ$+I$ classifications. 
Classes with either $PT$ or $PC$ are called real AZ$+I$ classes. There are eight real AZ$+I$ classes; we again label them by an integer $s$ (mod $8$). It is in these real classes where the difference between AZ and AZ$+I$ classifications appear. The goal of the paper is to clarify the bulk-boundary correspondence of the real AZ$+I$ classes through the entanglement spectrum. 

\subsection{Classification of gapped topological phases}
Classification of stable gapped topological phases for free fermions is known for both real AZ symmetry classes~\cite{PhysRevB.78.195125,kitaev2009periodic_2,ryu2010topological} and real AZ$+I$ symmetry classes~\cite{PhysRevB.82.115120,PhysRevB.96.155105}. In Table~\ref{tab:Periodic_table}, we provide the classification table up to three dimensions.
Derivation of the classification table for AZ classes goes back to Ref.~[\onlinecite{kitaev2009periodic_2}]. The classification table for real AZ$+I$ classes can be formally obtained as a special case of the classification table in Ref. [\onlinecite{PhysRevB.82.115120}]. In the terminology of Ref.[\onlinecite{PhysRevB.82.115120}], the classification of the $d_{\text{AZ}+I}$-dimensional Hamiltonian of the real AZ$+I$ class is formally equivalent to the classification of the Hamiltonian of the real AZ class which has $0$ momentum variable ($d=0$) and $d_{\text{AZ}+I}$ position variables ($D=d_{\text{AZ}+I}$).
\footnote{Note that in realistic topological insulators with lattice defects, $d<D$ cannot be realized. This is purely a discussion on the classification of topological phases.} 
The real AZ$+I$ class can also be considered as a complex AZ class with an additional order-two antiunitary symmetry, which is a special case of Sec.~III.B of Ref.[\onlinecite{PhysRevB.90.165114}].

One can notice that there is a structure, or periodicity, in the classification table; for AZ symmetry classes, same groups align along diagonal direction, whereas for AZ$+I$ symmetry classes, anti-diagonal alignment appears. Since we will make use of this periodicity upon deriving the bulk-boundary correspondence, we explain this periodicity in more detail below.

\begin{table*}[!]
\begin{center}
\caption{Topological classification tables for gapped systems for real AZ and real AZ$+I$ classes}
\begin{tabular}{c|c|cccc}
     s & real AZ & $d=0$ & $d=1$ & $d=2$ & $d=3$  \\  \hline \hline
     0 &   AI & $\mathbb{Z}$ & 0 & 0 & 0  \\
     1 &  BDI & $\mathbb{Z}_2$ & $\mathbb{Z}$ & 0 & 0  \\
     2 &    D & $\mathbb{Z}_2$ & $\mathbb{Z}_2$ & $\mathbb{Z}$ & 0  \\
     3 & DIII & 0 & $\mathbb{Z}_2$ & $\mathbb{Z}_2$ & $\mathbb{Z}$  \\
     4 &  AII & $2\mathbb{Z}$ & 0 & $\mathbb{Z}_2$ & $\mathbb{Z}_2$  \\
     5 &  CII & 0 & $2\mathbb{Z}$ & 0 & $\mathbb{Z}_2$  \\
     6 &    C & 0 & 0 & $2\mathbb{Z}$ & 0  \\
     7 &   CI & 0 & 0 & 0 & $2\mathbb{Z}$  \\
\end{tabular}
\quad \quad
\begin{tabular}{c|c|cccc}
    s & real AZ$+I$ & $d=0$ & $d=1$ & $d=2$ & $d=3$  \\  \hline \hline
    0 &   AI$^{\prime}$ & $\mathbb{Z}$ & $\mathbb{Z}_2$ & $\mathbb{Z}_2$ & 0  \\
    1 &  BDI$^{\prime}$ & $\mathbb{Z}_2$ & $\mathbb{Z}_2$ & 0 & $2\mathbb{Z}$  \\
    2 &    D$^{\prime}$ & $\mathbb{Z}_2$ & 0 & $2\mathbb{Z}$ & 0  \\
    3 & DIII$^{\prime}$ & 0 & $2\mathbb{Z}$ & 0 & 0  \\
    4 &  AII$^{\prime}$ & $2\mathbb{Z}$ & 0 & 0 & 0  \\
    5 &  CII$^{\prime}$ & 0 & 0 & 0 & $\mathbb{Z}$  \\
    6 &    C$^{\prime}$ & 0 & 0 & $\mathbb{Z}$ & $\mathbb{Z}_2$  \\
    7 &   CI$^{\prime}$ & 0 & $\mathbb{Z}$ & $\mathbb{Z}_2$ & $\mathbb{Z}_2$  \\
\end{tabular}
\label{tab:Periodic_table}
\end{center}
\end{table*}

The periodicity of the classification table can be obtained following the method presented in Ref.[\onlinecite{PhysRevB.82.115120}], which we now briefly review, for both real AZ and real AZ$+I$ classifications.
The Bloch Hamiltonian of a $d$-dimensional periodic lattice system is defined on a base space spanned by momentum $\mathbf{k}$, which takes values in a $d$-dimensional Brillouin zone $T^d$. As in Refs.[\onlinecite{kitaev2009periodic_2}] and [\onlinecite{PhysRevB.82.115120}], we will simplify the topological classification by treating the base space as a sphere $S^d$. This simplification is equivalent to focusing on the strong topological phases intrinsic to $d$-dimensional systems and ignoring the so-called weak topological phases, which originate from the lower dimensional topology.

In this section, we also focus on stable equivalence, that is, to study topological phases robust against adding trivial energy bands.
Focusing on stable equivalence corresponds to ignoring topological phases called fragile\cite{PhysRevLett.121.126402} and delicate\cite{PhysRevLett.126.216404} topological phases, which become trivial by adding trivial bands.
Later in Sec.~\ref{sec:fragile}, we extend our analysis of the bulk-edge correspondence to the cases of fragile topological phases. 

We denote the equivalence class of Hamiltonians that are stably equivalent to $H$ by $[H]$. 
In the following, without loss of generality, we assume that the Hamiltonian has an energy gap at the Fermi energy set at $E=0$.
The addition of two equivalence classes is defined as $[H_1]+[H_2]\equiv [H_1\oplus H_2]$, where $\oplus$ means the direct sum of the matrices. 
The additive inverse for $[H]$ is given by $[-H]$, and $[H\oplus(-H)]$ is always the identity element $[0]$ representing a trivial Hamiltonian, which follows from the observation that the sum of occupied and unoccupied bands yields trivial bands.
Subtraction is defined as $[H_1]-[H_2]\equiv[H_1\oplus(-H_2)]$. Then, the stable equivalent classes of the Hamiltonian form an abelian group called the $K$-group\cite{PhysRevB.82.115120,PhysRevB.90.165114}. The $K$-group describes the classification of stable topological phases. For a more detailed discussion of $K$-group, see Ref.[\onlinecite{PhysRevB.95.235425}].

We now show the periodicity in the classification table, which connects $K$-groups in higher dimensions to those in lower dimensions. 
First, we introduce dimensional raising map that sends a $d$-dimensional Hamiltonian $H^{(d)}(\mathbf{k})$ to a $(d+1)$-dimensional Hamiltonian $H^{(d+1)}(\mathbf{k},\theta)$. This new Hamiltonian depends on $\mathbf{k}$ $\in S^d$ as well as an additional variable $\theta$ defined in 0 $\le\theta\le\pi$. This is valid for both AZ and AZ$+I$ classifications. 
The definition of the dimensional raising map differs between Hamiltonians with chiral symmetry ($H^{(d)}_{\mathrm{c}}$) and Hamiltonians without chiral symmetry ($H^{(d)}_{\mathrm{nc}}$). 
Letting $\Gamma$ be the chiral symmetry operation of $H^{(d)}_{\mathrm{c}}$, the dimensional raising map for $H^{(d)}_{\mathrm{c}}$ is defined as 
\begin{align}
H^{(d+1)}_{\mathrm{nc}}(\mathbf{k},\theta)
&=\sin\theta H^{(d)}_{\mathrm{c}}(\mathbf{k})+\cos\theta \Gamma,\label{Hnc_d+1}
\end{align}
and that for $H^{(d)}_{\mathrm{nc}}$ is defined as
\begin{align}
H^{(d+1)}_{\mathrm{c}}(\mathbf{k},\theta)&=\sin\theta H^{(d)}_{\mathrm{nc}}(\mathbf{k})\otimes\tau_z+\cos\theta\mathbb{I}\otimes\tau_a,
\label{Hc_d+1}
\end{align}
where $\tau_i$ ($i=x,y,z$) are Pauli matrices. In Eq.~(\ref{Hc_d+1}), $\tau_a=\tau_x$ or $\tau_y$ is chosen to preserve the symmetry of $H^{(d)}_{\mathrm{nc}}(\mathbf{k})$. 
At $\theta=0$ and $\theta=\pi$, the mapped Hamiltonians $H^{(d+1)}_{\mathrm{nc}}(\mathbf{k},\theta)$ and $H^{(d+1)}_{\mathrm{c}}(\mathbf{k},\theta)$ are independent of $\mathbf{k}$. This means that the base space of the mapped Hamiltonian $(\mathbf{k},\theta)\in S^d\times[0,\pi]$ can be identified as the $suspension$ $S S^d=S^{d+1}$, by contracting $S^d\times\{0\}$ to one point and similarly $S^d\times\{\pi\}$ to another point.  Thus the new Hamiltonian is naturally defined on a $(d+1)$-dimensional sphere $S^{d+1}$. 

The dimensional raising maps eliminate or add the chiral symmetry, and shifts the AZ or AZ$+I$ class by one. For the complex AZ and AZ$+I$ classes, if the original Hamiltonian belongs to class AIII, i.e., has only a chiral symmetry, the mapped Hamiltonian has no symmetry, i.e., belongs to class A. On the other hand, if the original Hamiltonian has no symmetry and belongs to class A, the mapped Hamiltonian has chiral symmetry $\Gamma=\mathbb{I}\otimes i\tau_z\tau_a$ and belongs to class AIII. For the real AZ or AZ$+I$ classes, carefully following the change of symmetry classes as shown in Appendix \ref{class shift}, one can show that the real AZ class shifts from $s$ to $s+1$, and the real AZ$+I$ class shifts from $s$ to $s-1$ (mod $8$). 

In summary, Eqs.~(\ref{Hnc_d+1}) and (\ref{Hc_d+1}) define a map that sends a $d$-dimensional Hamiltonian $H^{(d)}(\mathbf{k})$ with class $s$ to a $(d+1)$-dimensional Hamiltonian $H^{(d+1)}(\mathbf{k},\theta)$ with class $(s+1)$ or $(s-1)$. Since for any $(\mathbf{k}, \theta) \in S^d \times [0,\pi]$ the energy gaps of the mapped Hamiltonians $H^{(d+1)}_{\mathrm{nc}}(\mathbf{k},\theta)$ and $H^{(d+1)}_{\mathrm{c}}(\mathbf{k},\theta)$ remain open, two Hamiltonians that are topologically equivalent are mapped to topologically equivalent Hamiltonians. In addition, the dimensional raising map preserves the direct sum structure of the Hamiltonian. Therefore, the dimensional raising map leads to a homomorphism from the lower dimensional $K$ group to the higher dimensional $K$ group:
\begin{align}
K_{\mathrm{TI}}^{\mathrm{AZ}}(s,d) &\longrightarrow K_{\mathrm{TI}}^{\mathrm{AZ}}(s+1,d+1),
\label{K_AZ_d_arrow_d+1}
\\
K_{\mathrm{TI}}^{\mathrm{AZ}+I}(s,d) &\longrightarrow K_{\mathrm{TI}}^{\mathrm{AZ}+I}(s-1,d+1).
\label{K_AZ+I_d_arrow_d+1}
\end{align}
Here, $K_{\mathrm{TI}}^{\mathrm{AZ}(+I)}(s,d)$ is a $K$ group that classifies the class $s$ gapped Hamiltonian defined on the $d$-dimensional sphere $S^d$ under the AZ($+I$) classification.
Furthermore, the homomorphism in Eq.~(\ref{K_AZ_d_arrow_d+1}) for AZ classification has an inverse as shown in Ref.[\onlinecite{PhysRevB.82.115120}], giving an isomorphism between the $K$ groups:
\begin{align}
K_{\mathrm{TI}}^{\mathrm{AZ}}(s,d) &= K_{\mathrm{TI}}^{\mathrm{AZ}}(s+1,d+1).
\label{K_AZ_d-d+1}
\end{align}
Note that the dimensional-raising map itself generally does not have an inverse map. However, any $(d+1)$-dimensional Hamiltonian can be continuously transformed into an image of the dimensional-raising map using stable equivalence\cite{PhysRevB.82.115120}. 

Similarly, the inverse can also be constructed for the homomorphism in Eq.~(\ref{K_AZ+I_d_arrow_d+1}) for AZ$+I$ classification, by following the formalism for the case of position variables presented in Ref.~[\onlinecite{PhysRevB.82.115120}]. 
This leads to the isomorphism between the $K$ groups: 
\begin{align}
K_{\mathrm{TI}}^{\mathrm{AZ}+I}(s,d) &= K_{\mathrm{TI}}^{\mathrm{AZ}+I}(s-1,d+1).
\label{K_AZ+I_d-d+1}
\end{align}

The relations Eq.~(\ref{K_AZ_d-d+1}) and Eq.~(\ref{K_AZ+I_d-d+1}) are exactly the diagonal and anti-diagonal periodicities present in the classification table for real AZ class and real AZ$+I$ class, respectively.

\subsection{Stable Fermi surfaces}
In order to relate bulk topology with boundary physics, we should also obtain classification of stable gapless modes, or stable zero-energy states. Since Fermi surface is a subset of $\mathbf{k}$-space consisting of zero-energy states, the classification of stable zero-energy states is nothing but the classification of stable Fermi surface. 
The stability of the Fermi surface can also be understood as a consequence of topology~\cite{PhysRevLett.95.016405,PhysRevLett.110.240404,PhysRevB.89.075111}.

For the AZ classification, it is known that the following equality holds between the group that classifies stable gapped topological phases and the group that classifies stable Fermi surfaces\cite{PhysRevB.89.075111}:
\begin{align}
K_{\mathrm{TI}}^{\mathrm{AZ}}(s,d)
=K_{\mathrm{FS}}^{\mathrm{AZ}}(s+1,d). 
\label{K_TI_FS_AZ}
\end{align}
Here, $K_{\mathrm{FS}}^{\mathrm{AZ}}(s+1,d)$ is the group that classifies the class $(s+1)$ Fermi surface with codimension $d$. 
Since the subset with codimension $d$ in $d$-dimensional space corresponds to points, $K_{\mathrm{FS}}^{\mathrm{AZ}}(s+1,d)$ gives the classification of point nodes in $d$-dimensional space with the symmetry class $(s+1)$. 

For the AZ$+I$ classification, since $PT$ and $PC$ do not change $\mathbf{k}$, the Hamiltonian restricted to $S^{d-1}$ surrounding the band degeneracy belongs to the gapped Hamiltonian of AZ$+I$ classification in the same symmetry class. Therefore, the following equation holds:
\begin{align}
K_{\mathrm{TI}}^{\mathrm{AZ}+I}(s,d-1)
=K_{\mathrm{FS}}^{\mathrm{AZ}+I}(s,d). 
\label{K_TI_FS_AZ+I_d-1}
\end{align}
By combining Eqs.~(\ref{K_AZ+I_d-d+1}) and (\ref{K_TI_FS_AZ+I_d-1}), we obtain
\begin{align}
K_{\mathrm{TI}}^{\mathrm{AZ}+I}(s,d)
=K_{\mathrm{FS}}^{\mathrm{AZ}+I}(s+1,d),
\label{K_TI_FS_AZ+I}
\end{align}
which is exactly the same as the relation for the AZ classification Eq.~\eqref{K_TI_FS_AZ}.

We have thus obtained $K$-groups and relations between them for bulk Hamiltonians and Fermi surfaces for both AZ and AZ$+I$ symmetry classes. We will now move on to discuss the bulk-boundary correspondence.

\section{Bulk-boundary correspondence}
\label{sec:bbc}

\subsection{AZ classes}

We start from giving a short proof of the bulk-boundary correspondence in AZ symmetry classes from the $K$-group perspective~\cite{PhysRevB.89.075111}. 
Let us consider a gapped $d$-dimensional system in the symmetry class $s$ in the real AZ classification. The topology of such Hamiltonians are given by the $K$-group $K_{\mathrm{TI}}^{\mathrm{AZ}}(s,d)$.
We then consider imposing an open boundary condition along one direction, but keep the $d-1$ directions to be periodic (or infinitely long). The system can then be described by a Hamiltonian parametrized by a momentum $\mathbf{k}_\perp$, which is a $d-1$ dimensional momentum perpendicular to the direction where the open boundary condition is taken. (In other words, the momentum $\mathbf{k}_\perp$ is parallel to the edge.) 
If the length along the direction where the open boundary condition is applied is sufficiently long, the eigenstates localized at both ends will hardly hybridize with each other and can be described by independent effective Hamiltonians. 
Let us call one such effective Hamiltonian, consisting of states localized at one of these edges and parametrized by $\mathbf{k}_\perp$, the edge Hamiltonian.
Since the symmetries $T$ and $C$ are local, the edge Hamiltonian and the bulk Hamiltonian belong to the same symmetry class $s$. Since the edge Hamiltonian is $d-1$ dimensional, the gapless modes of the edge Hamiltonian is classified by the $K$-group $K_{\mathrm{FS}}^{\mathrm{AZ}}(s,d-1)$.

On the other hand, by combining Eqs.~(\ref{K_AZ_d-d+1}) and (\ref{K_TI_FS_AZ}), the following relation holds:
\begin{align}
K_{\mathrm{TI}}^{\mathrm{AZ}}(s,d)
=K_{\mathrm{FS}}^{\mathrm{AZ}}(s,d-1),
\label{AZ_BBC}
\end{align}
which means that the bulk gapped topological phases and the gapless edge states share the same topological classification. This is nothing but the statement of the bulk-boundary correspondence.

For practical purposes, it would also be useful to concretely construct the topological invariant for the bulk Hamiltonian, which corresponds to an element in $K_{\mathrm{TI}}^{\mathrm{AZ}}(s,d)$, and similarly the topological invariant of the edge Hamiltonian, which corresponds to an element in $K_{\mathrm{FS}}^{\mathrm{AZ}}(s,d-1)$. (For example, in class A, one wants to find that the Chern number is equal to the number of chiral edge modes.) For this purpose, one can first construct a representative Hamiltonian that is the generator of $K_{\mathrm{TI}}^{\mathrm{AZ}}(s,d)$, and show that the corresponding edge Hamiltonian is a generator of $K_{\mathrm{FS}}^{\mathrm{AZ}}(s,d-1)$. General cases can be shown by considering the direct sum of the generator Hamiltonians. Using Dirac matrices one can obtain such a construction~\cite{ryu2010topological,PhysRevB.89.075111}.

\subsection{AZ$+I$ classes}

Unlike AZ classification, the symmetries of AZ$+I$ classifications such as $PT$ symmetry and $PC$ symmetry, are position dependent. Therefore, the edge Hamiltonian does not have these symmetries and, in fact, does not belong to any real AZ$+I$ class in general.
As a consequence, the ordinary bulk-boundary correspondence does not hold, and the edge spectrum is generally gapped in the AZ$+I$ symmetry classes.

Although the ordinary bulk-boundary correspondence does not hold, we can still obtain an analog of Eq.~(\ref{AZ_BBC}) for AZ$+I$ classification by combining Eqs.~(\ref{K_AZ+I_d-d+1}) and (\ref{K_TI_FS_AZ+I}):
\begin{align}
K_{\mathrm{TI}}^{\mathrm{AZ}+I}(s,d)
=K_{\mathrm{FS}}^{\mathrm{AZ}+I}(s+2,d-1). 
\label{AZ+I_BBC}
\end{align}
This is one of the key results of this paper.
What we are going to show below is that, for a gapped bulk Hamiltonian which belongs to class $s$ in the real AZ$+I$ classification at $d$ dimensions, the corresponding stable gapless {\it entanglement spectrum} belongs to class $s+2$ in the real AZ$+I$ classification at $d-1$ dimensions, described by the right-hand side of Eq.~(\ref{AZ+I_BBC}). Equation~(\ref{AZ+I_BBC}) thus describes a variant of the bulk-boundary correspondence for the real AZ$+I$ classification, which we call {\it bulk-entanglement spectrum correspondence }.

\subsection{Single-particle entanglement spectrum for general finite systems}
\label{sec:spes_general}

From here, we will discuss the single-particle entanglement spectrum~\cite{Ingo_Peschel_2003,peschel2012entanglement}. For the convenience of the reader, we will start with the many-body entanglement spectrum for non-interacting systems. 
The many-body entanglement spectrum was originally proposed as a tool for identifying topological order\cite{PhysRevLett.101.010504,PhysRevB.81.064439}.
We consider partitioning the total Hilbert space, $\mathcal{H}_{\text{tot}}$, into two parts: $\mathcal{H}_{\text{tot}} = \mathcal{H}_L \otimes \mathcal{H}_R$. The ground state, $\ket{\Psi_{\text{GS}}}$, can then be expressed in terms of the Schmidt decomposition:
\begin{align}
\ket{\Psi_{\text{GS}}} = \sum_j \lambda_j \ket{\psi_L^{(j)}} \otimes \ket{\psi_R^{(j)}}.
\end{align}
The many-body entanglement spectrum is defined by the set $\{|\lambda_j|^2\}$. This set is equivalent to the eigenvalues of the reduced density matrix, $\rho_L$, which is given by:
\begin{align}
\rho_L &= \text{Tr}_R(\ket{\Psi_{\text{GS}}}\bra{\Psi_{\text{GS}}})= \sum_j |\lambda_j|^2 \ket{\psi_L^{(j)}}\bra{\psi_L^{(j)}}.
\end{align}

The above definition of the many-body entanglement spectrum can be applied to any many-body state. For non-interacting fermions, as we show, we can derive a simple expression of the many-body entanglement spectrum in terms of the correlation functions\cite{Ingo_Peschel_2003}, known as the single-particle entanglement spectrum\cite{PhysRevLett.104.130502,PhysRevB.82.241102,PhysRevB.83.245132,PhysRevB.84.195103,PhysRevB.85.165120,PhysRevB.108.075129}. 
In this subsection, following Ref.~[\onlinecite{PhysRevB.108.075129}], we briefly introduce the single-particle entanglement spectrum and an emergent antisymmetry.

First, we introduce the single-particle entanglement spectrum for general finite systems. 
We consider the following Hamiltonian:
\begin{align}
    \hat{H}=\sum_{i,j\in D}h_{i,j}\hat{c}_i^{\dagger}\hat{c}_j.
\end{align}
Here, $i,j\in D$ are some degrees of freedom of the finite system, such as lattice sites, orbits, and spins, and $\hat{c}_i^\dagger$ ($\hat{c}_i$) are the creation (annihilation) operator of a particle in the $i$-th degree of freedom. The elements of the Hamiltonian $h_{i,j}$ should obey the Hermiticity condition $h_{i,j} = h_{j,i}^*$. We suppose that $D$ is a disjoint union of $A$ and $B$, that is, any element of $D$ can be divided into either $A$ or $B$. 
We want to construct a single-particle correlation function from the Hamiltonian $\hat{H}$. For this purpose, let us denote by $H$ the Hamiltonian matrix whose $(i,j)$-component is $h_{i,j}$, and denote its $n$-th eigenvalue and the corresponding eigenvector by $E_n$ and $|\Phi^{(n)}\rangle$. The eigenvector $|\Phi^{(n)}\rangle$ is a multi-component vector whose length corresponds to the total degrees of freedom in the system; we denote $i$-th component of $|\Phi^{(n)}\rangle$ by $\Phi_i^{(n)}$.
The many-body ground state of non-interacting fermions is a single Slater determinant of all single-particle eigenstates of $\hat{H}$ below the Fermi energy $E_\text{F} = 0$, which can be written as 
\begin{align}
    \ket{\text{GS}}=\prod_{E_n<0}\sum_i \hat{c}^{\dagger}_i\Phi_i^{(n)} \ket{0},
\end{align}
where $|0\rangle$ is the vacuum state.
The correlation function of noninteracting system is then defined by
\begin{align}
 C^{\text{cor}}_{i,j}
 =
 \bra{\text{GS}}\hat{c}_j^{\dagger} \hat{c}_i\ket{\text{GS}}
 =
 \sum_{E_n<0} \Phi^{(n)}_{i} \Phi^{(n) *}_{j},
\end{align}
which can also be written as
\begin{align}
     C^{\text{cor}}=\sum_{E_n<0}|\Phi^{(n)}\rangle
    \langle\Phi^{(n)}|.
\end{align}
Since $D$ is the disjoint union of $A$ and $B$, the correlation matrix $C^{\text{cor}}$, whose $(i,j)$-component is $C^{\text{cor}}_{i,j}$, can be written in the following form:
\begin{align}
    C^{\text{cor}}=\begin{pmatrix}
        C_A & C_{AB} \\
        C_{BA} & C_B
    \end{pmatrix},
    \label{C_block}
\end{align}
where $C_A$ is a matrix made of $C^{\text{cor}}_{i,j}$ where $i$ and $j$ run only within the region $A$, namely $C_A\equiv(C^{\text{cor}}_{i,j})|_{i,j\in A}$. Similarly, $C_B\equiv(C^{\text{cor}}_{i,j})|_{i,j\in B}$, $C_{AB}\equiv(C^{\text{cor}}_{i,j})|_{i\in A, j \in B}$, and $C_{BA}\equiv(C^{\text{cor}}_{i,j})|_{i \in B, j\in A}$. It can be shown that the eigenvalues of $C^{\text{cor}}$ are 0 and 1, and the eigenvalues of $C_A$ and $C_B$ are in the range $[0, 1]$.
\footnote{Since $(C^{\text{cor}})^2=C^{\text{cor}}$, we have $C_A^2+C_{AB}C_{BA}=C_A$. From the Hermiticity of $C^{\text{cor}}$, $C_{AB}=C_{BA}^{\dagger}$. Consequently, $C_{BA}^{\dagger}C_{BA}=C_A(\mathbb{I}-C_A)$ holds. Since the left-hand side is positive-semidefinite, the eigenvalue $\xi_l$ of $C_A$ must satisfy $\xi_l(1-\xi_l)\geq0$. This condition is equivalent to $0\leq\xi_l\leq1$. A similar argument can be applied to $C_B$.}
Furthermore, the eigenvalues of $C_A$ and $(\mathbb{I}-C_B)$, excluding 0 and 1, match exactly, including their multiplicities.
\footnote{
From Eq.~(\ref{B3}), for the eigenstate $|f_l^{A}\rangle$ of $C_A$ with the eigenvalue $\xi_l$, the state $C_{BA}|f_l^{A}\rangle$ is the eigenstate of $(\mathbb{I}-C_B)$ with the eigenvalue $\xi_l$. 
Furthermore, from Eq.~(\ref{B4}), if $\xi_l\neq0,1$, the magnitude of state $C_{BA}|f_l^{A}\rangle$ is not 0. Therefore, the eigenvalues of $C_A$ and $(\mathbb{I}-C_B)$, excluding 0 and 1, match exactly. 
}
Let us denote the eigenvalues of $C_A$ (or equivalently $\mathbb{I}-C_B$) whose values are not equal to 0 or 1 by $\xi_l$. Let us denote the number of such eigenvalues different from 0 or 1 by $N$.

As we detail in the Appendix.~\ref{Deriv_GS_vac}, the ground state of the original Hamiltonian $\hat{H}$ can be written in the following form in the Schmidt decomposition:
\begin{align}
\ket{\text{GS}}
=&\sum_{\{n_l\}}(-1)^{S(\{n_l\})}
\prod_l\xi_l^{\frac{n_l}{2}}
(1-\xi_l)^{\frac{1-n_l}{2}}
\ket{\alpha_{\{n_l\}}^{A}}\otimes\ket{\alpha_{\{n_l\}}^{B}},  \label{eq:gs}
\\
\ket{\alpha_{\{n_l\}}^{A}}&=
\prod_{n_A}\Bigl{(}\sum_{i\in A}(g^A_{n_A})_i c^{\dagger}_i\Bigr{)}\prod_l\Bigl{(}\sum_{i\in A}(f^A_l)_i c^{\dagger}_i\Bigr{)}^{n_l}\ket{\text{0}^{A}},
\\
\ket{\alpha_{\{n_l\}}^{B}}&=\prod_{n_B}\Bigl{(}\sum_{i\in B}(g^B_{n_B})_i c^{\dagger}_i\Bigr{)}\prod_l
\Bigl{(}\sum_{j\in B}(f^B_l)_j c^{\dagger}_j
\Bigr{)}^{1-n_l}\ket{\text{0}^{B}},
\end{align}
where $n_l \in \{ 0, 1\}$; the set $\{ n_l \}$ takes its value in $\{0, 1\}^N$and the summation over $\{n_l\}$ is for all possible values of $\{ n_l \} \in \{0, 1\}^N$. The factor $S(\{n_l\})$ is the sign factor that can arise from the commutation of creation operators. The state $|0^{A(B)}\rangle$ is the vacuum state of the region $A$ ($B$), $(g_{n_{A(B)}}^{A(B)})_i$ is the $i$-th component of the $n_{A(B)}$-th eigenvector of $C_{A(B)}$ with the eigenvalue 1, and $(f_l^{A(B)})_i$ is the $i$-th component of the eigenvector of $C_{A(B)}$ with the eigenvalue $\xi_l$. 

Therefore, the many-body entanglement spectrum of $\ket{\text{GS}}$ is
\begin{align}
\lambda_{\{n_l\}}=\prod_l\xi_l^{n_l}
(1-\xi_l)^{1-n_l}.
\end{align}
The many-body entanglement spectrum $\{\lambda_{\{n_l\}}\}$ is thus completely determined by $\{\xi_l\}$.

\begin{figure}[t]
\centerline{\includegraphics[width=8.5cm,clip]{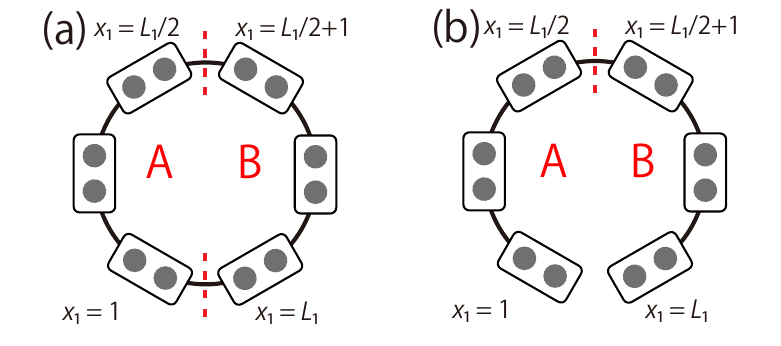}}
  \caption{Conceptual figure of finite lattice systems and entanglement cuts, featuring (a) periodic boundary condition and (b) open boundary condition along $x_1$-direction. Each circle represents a site, square boxes denote unit cells, and red dashed lines indicate entanglement cuts. Since there are two entanglement cuts in (a), approximately twice as many states contribute to the entanglement spectrum compared to (b). Thus, in a sense, the entanglement spectrum of (a) can be seen as 'double' of that of (b).
  }
    \label{entanglement_cut}
\end{figure}

The correlation matrix $C^{\text{cor}}$ and the flattened Hamiltonian $H_{\text{flat}}$ are also intimately related. The flattened Hamiltonian is the Hamiltonian obtained by taking the original Hamiltonian and deforming it so that the eigenvalues below 0 go to $-1/2$ and the eigenvalues above 0 go to $+1/2$.
The flattened Hamiltonian is related to the correlation function via:
\begin{align}
    H_{\text{flat}} = \frac{1}{2} \mathbb{I} - C^{\text{cor}},
    \label{H_flat_general}
\end{align}
where $\mathbb{I}$ is the identity matrix whose dimension equals to $C^{\text{cor}}$. 
The eigenvalues of the flattened Hamiltonian are thus $\pm 1/2$, and its restriction to region $A$, $H_{\text{flat},A}=(H_{\text{flat}})_{i,j}|_{i,j\in A}$, have eigenvalues in the range $[-1/2, +1/2]$.
The eigenvalues $\{\epsilon_l\}$ of $H_{\text{flat},A}$ are $1/2 - \xi_l$ or $\pm 1/2$. 
Using $\epsilon_l = 1/2 - \xi_l$, the many-body entanglement spectrum of $\ket{\text{GS}}$ can be written as
\begin{align}
\lambda_{\{n_l\}}=\prod_l(1/2-\epsilon_l)^{n_l}
(1/2+\epsilon_l)^{1-n_l}.
\label{MES_ES}
\end{align}
Thus, the eigenvalues $\{\xi_l\}$ of $C_A$ and the eigenvalues $\{\epsilon_l\}$ of $H_{\text{flat},A}$ contain the equivalent information when reconstructing the many-body entanglement spectrum.
(We note that we defined $\{ \xi_l \}$ not including $0$ and $1$, but we will include $\pm 1/2$ in $\{ \epsilon_l \}$ for notational convenience.)
In the subsequent discussions, unless otherwise specified, we call the spectrum $\{\epsilon_l\}$ of $H_{\text{flat},A}$ as the {\it single-particle entanglement spectrum}, or simply the entanglement spectrum.

\subsection{Single-particle entanglement spectrum for finite lattice systems with open boundary condition}

We now explain how to calculate the entanglement spectrum for a finite lattice system. In our previous study Ref.~[\onlinecite{PhysRevB.108.075129}], to calculate the entanglement spectrum, we considered a finite system with the periodic boundary condition imposed in all directions, as shown in Fig.~\ref{entanglement_cut}(a). 
\footnote{
The single-particle entanglement spectrum in this approach is equivalent to the open boundary spectrum of a Hamiltonian that features a flattened bulk spectrum.
}
In this paper, we choose to apply the open boundary condition in one direction, which we take to be $x_1$-direction, upon defining the entanglement spectrum; the situation is schematically shown in Fig.~\ref{entanglement_cut}(b)\cite{PhysRevB.84.195103}. 
Upon calculating the entanglement spectrum, we need to ``cut" the system in two. If we apply the periodic boundary condition in $x_1$-direction, we need to have two cuts, whereas if we apply the open boundary condition, we only need one cut (Fig.~\ref{entanglement_cut}). 
Both approaches, either periodic or open along $x_1$-direction, give equivalent results, but since the former (periodic) approach has two cuts rather than one cut for the latter (open), the entanglement spectrum is doubled in the former approach.
In the approach we take here, which is to apply the open boundary condition along $x_1$-direction, the resulting entanglement spectrum has contributions solely from one cut, which we find to be more convenient.

The single-particle entanglement spectrum for the finite lattice system with open boundary condition is calculated by the following process. First, let the Hamiltonian of the open boundary condition in the $x_1$-direction be $H(\mathbf{k}_{\perp})$, where $\mathbf{k}_{\perp}$ is the wave vector perpendicular to the $x_1$-direction, and find its eigenvalues and eigenstates. 
Now, take eigenstates whose eigenvalues are less than the Fermi energy (which is set to zero), and construct the projection $P_{\text{occ}}(\mathbf{k}_{\perp})$ onto the eigenspace spanned by these eigenstates. 
Then, we define the flattened Hamiltonian $H_{\mathrm{flat}}(\mathbf{k}_{\perp})$ by:
\begin{align}
    H_{\mathrm{flat}}(\mathbf{k}_{\perp})\equiv\frac{1}{2}\mathbb{I}-P_{\text{occ}}(\mathbf{k}_{\perp}).
\end{align}

We assume that, as schematically shown in  Fig.~\ref{entanglement_cut}(b), the lattice is divided into two parts with the equal length; part ``$A$", for $1 \le x_1 \le L_1/2$ and part ``$B$" for $1 + L_1/2 \le x_1 \le L_1$, where $L_1$, the number of unit cells in $x_1$ direction, is assumed to be an even integer. Here, the two parts are made the same length to preserve $PT$ and $PC$ symmetry, as will be discussed in the following subsection.
We express $H_{\mathrm{flat}}(\mathbf{k}_{\perp})$ in the following form:
\begin{align}
    H_{\mathrm{flat}}(\mathbf{k}_{\perp}) = \begin{pmatrix}
    H_{\mathrm{flat},A}(\mathbf{k}_{\perp}) &     H_{\mathrm{flat},AB}(\mathbf{k}_{\perp}) \\ 
    H_{\mathrm{flat},BA}(\mathbf{k}_{\perp}) &      H_{\mathrm{flat},B}(\mathbf{k}_{\perp}) 
        \end{pmatrix}, \label{eq:x1obc}
\end{align}
where, for instance, $H_{\mathrm{flat},AB}(\mathbf{k}_{\perp})$ is a matrix formed by restricting $[H_{\mathrm{flat}}(\mathbf{k}_{\perp})]_{x_1\alpha,x_1^\prime\beta}$ to $1 \le x_1 \le L_1/2$ and $1 + L_1/2 \le x_1^\prime \le L_1$.
The eigenvalues of $H_{\mathrm{flat},A}(\mathbf{k}_{\perp})$ provide the single-particle entanglement spectrum of the region $A$. 

As we show in the Appendix~\ref{General_Xi}, an operator $\Xi_{BA}$ defined below provides an important antisymmetry:
\begin{align}
    \Xi_{BA} &\equiv \sum_{n \text{ for }\epsilon_n\neq\pm1/2} |\phi_n \rangle\langle \psi_n|,
    \label{Xi_ep_psi_OBC}
    \\
    |\phi_n \rangle&\equiv-\frac{H_{\mathrm{flat},BA}(\mathbf{k}_{\perp})}{\sqrt{ 1/4 - \epsilon_n^2}}|\psi_n \rangle,
\end{align} 
where $|\psi_n \rangle$ are the normalized eigenstates of $H_{\mathrm{flat},A}(\mathbf{k}_{\perp})$ with the eigenvalues $\epsilon_n\neq\pm1/2$. 
This operator defines an ``emergent antisymmetry" because the relation $H_{\mathrm{flat},B}(\mathbf{k}_{\perp})\Xi_{BA}\ket{\psi_n}=-\epsilon_n\Xi_{BA}\ket{\psi_n}$ holds.
We note that $\ket{\phi_n}=\Xi_{BA}\ket{\psi_n}$ is a normalized eigenstate of $H_{\mathrm{flat},B}(\mathbf{k}_{\perp})$.
Thus, the operator $\Xi_{BA}$ acting on an eigenstate of $H_{\mathrm{flat},A}(\mathbf{k}_{\perp})$ gives an eigenstate of $H_{\mathrm{flat},B}(\mathbf{k}_{\perp})$ with the opposite eigenvalue. It is useful to define an operator $\Xi$ which acts on the entire Hilbert space of both regions $A$ and $B$ by
\begin{align}
    \Xi \equiv \begin{pmatrix} 0 & \Xi_{AB} \\ \Xi_{BA} & 0\end{pmatrix},
\end{align}
where $\Xi_{AB} \equiv \Xi_{BA}^\dagger$ is an operator which transforms an eigenstate of $H_{\mathrm{flat},B}(\mathbf{k}_{\perp})$ into that of $H_{\mathrm{flat},A}(\mathbf{k}_{\perp})$ with the opposite eigenvalue.
As we show in the next subsection, the composition of the $PT$ and $PC$ symmetries and the operator $\Xi_{BA}$ gives the emergent symmetry of $H_{\mathrm{flat},A}$, which determines the effective AZ$+I$ class of the entanglement spectrum.

\subsection{Effective AZ$+I$ class of the entanglement spectrum}

In this subsection, we discuss symmetry properties of $H_{\mathrm{flat},A}(\mathbf{k}_{\perp})$. Since we will be interested in eigenvalues $\epsilon_n \neq \pm 1/2$, in this subsection we restrict the domain of $H_{\mathrm{flat},A}(\mathbf{k}_{\perp})$ to its eigenspace with eigenvalues $\epsilon_n \neq \pm 1/2$.~\footnote{What we mean by this restriction is to consider $P_{\epsilon_n \neq \pm 1/2} H_{\mathrm{flat},A}(\mathbf{k}_{\perp}) P_{\epsilon_n \neq \pm 1/2}$, where $P_{\epsilon_n \neq \pm 1/2}$ is the projection operator onto the eigenspace of $H_{\mathrm{flat},A}(\mathbf{k}_{\perp})$ with $\epsilon_n \neq \pm 1/2$}
We first note that $PT$ and $PC$ symmetries present in the original Hamiltonian, with the entanglement cut at the center of inversion, swap subsystems $A$ and $B$.
The $PT$ and $PC$ operators can then be written in the following block-diagonal forms:
\begin{align}
    PT = \begin{pmatrix} 0 & (PT)_{AB} \\ (PT)_{BA} & 0\end{pmatrix},
    PC = \begin{pmatrix} 0 & (PC)_{AB} \\ (PC)_{BA} & 0\end{pmatrix}.
\end{align}
The off-diagonal parts $(PT)_{BA}$ and $(PC)_{BA}$ transform the eigenstates of $H_{\mathrm{flat},A}(\mathbf{k}_{\perp})$ into those of $H_{\mathrm{flat},B}(\mathbf{k}_{\perp})$ (for more details, see Appendix \ref{commutation relation}). 
The properties of $PT$ and $\Xi$ imply that their composite operations $PT\Xi$ and $PC\Xi$, restricted to region $A$ ($B$), transfer eigenstates of $H_{\mathrm{flat},A (B)}(\mathbf{k}_{\perp})$ into the eigenstates of itself. Since $PT$ commutes and $\Xi$ anti-commutes with the Hamiltonian, the product $PT\Xi$ restricted to region $A$ ($B$) is the anti-symmetry of $H_{\mathrm{flat},A (B)}(\mathbf{k}_{\perp})$. Similarly, $PC\Xi$ is the symmetry of  $H_{\mathrm{flat},A (B)}(\mathbf{k}_{\perp})$. Furthermore, since $P$ and $\Xi$ are linear operators, and $T$ and $C$ are anti-linear operators, the composite operations $PT\Xi$ and $PC\Xi$ are anti-linear operators. In summary, $PT\Xi$ behave in a $PC$-like manner, and $PC\Xi$ behave in a $PT$-like manner, respectively.

As detailed in Appendix \ref{commutation relation}, $\Xi$ commutes and anti-commutes with $PT$ and $PC$ respectively, satisfying $[PT,\Xi]=0$ and $\{PC,\Xi\}=0$. Given that $\Xi^2$ behaves as an identity matrix for the states with $\epsilon_n (\neq\pm1/2$), it follows that $(PT\Xi)^2=(PT)^2$ and $(PC\Xi)^2=-(PC)^2$. As a result, the AZ$+I$ class of $H_{\mathrm{flat},A}(\mathbf{k}_{\perp})$ is uniquely determined from the AZ$+I$ class of the original Hamiltonian. By calculating the squared values of $PT\Xi$ and $PC\Xi$ for each class, we find how the AZ$+I$ symmetry class of $H_{\mathrm{flat},A}(\mathbf{k}_{\perp})$ shifts from that of the original Hamiltonian, as summarized in Table \ref{tab:Bulk_edge_AZ+I}. In summary, when the AZ$+I$ class of the original bulk Hamiltonian is represented by $s$ (mod 8), that of $H_{\mathrm{flat},A}(\mathbf{k}_{\perp})$ is $(s+2)$ (mod 8).

\begin{table}[!]
\begin{center}
\caption{AZ$+I$ symmetry classes of bulk and effective edge Hamiltonian of the entanglement spectrum.The order of the third and fourth columns is based on the fact that $PC\Xi$ and $PT\Xi$ are symmetric and anti-symmetric, respectively, and behave in a $PT$-like and $PC$-like manner, respectively.}
\begin{tabular}{cc|cc|c||ccc}
 $PT$ & $PC$ & $PC\Xi$ & $PT\Xi$ & $\Gamma$ & Bulk class & & ES class \\  \hline \hline
 $+$ & 0  & 0 & $+$ & 0 & AI$^{\prime}$ & $\rightarrow$ & D$^{\prime}$ \\
 $+$ & $+$  & $-$ & $+$ & 1 & BDI$^{\prime}$ & $\rightarrow$ & DIII$^{\prime}$ \\
 0 & $+$  & $-$ & 0 & 0 & D$^{\prime}$ & $\rightarrow$ & AII$^{\prime}$ \\
 $-$ & $+$  & $-$ & $-$ & 1 & DIII$^{\prime}$ & $\rightarrow$ & CII$^{\prime}$ \\
 $-$ & 0  & 0 & $-$ & 0 & AII$^{\prime}$ & $\rightarrow$ & C$^{\prime}$ \\
 $-$ & $-$  & $+$ & $-$ & 1 & CII$^{\prime}$ & $\rightarrow$ & CI$^{\prime}$ \\
 0 & $-$  & $+$ & 0 & 0 & C$^{\prime}$ & $\rightarrow$ & AI$^{\prime}$ \\
 $+$ & $-$  & $+$ & $+$ & 1 & CI$^{\prime}$ & $\rightarrow$ & BDI$^{\prime}$ \\
\end{tabular}
\label{tab:Bulk_edge_AZ+I}
\end{center}
\end{table}

\subsection{Bulk-boundary correspondence in the entanglement spectrum}
Finally, we show a variant of bulk-boundary correspondence which holds between the bulk topology and the entanglement spectrum. 
There exists a relation between the $d$ dimensional class $s$ gapped Hamiltonians in AZ$+I$ classification and $d-1$ dimensional class $s+2$ gapless Hamiltonians in AZ$+I$ classification as derived in Eq.~(\ref{AZ+I_BBC}), which we rewrite here:
\begin{align}
K_{\mathrm{TI}}^{\mathrm{AZ}+I}(s,d)
=K_{\mathrm{FS}}^{\mathrm{AZ}+I}(s+2,d-1). 
\end{align}
As shown in the previous subsection, within the framework of the AZ$+I$ classification, for a bulk Hamiltonian belonging to a $d$-dimensional class $s$, 
the single-particle entanglement Hamiltonian $H_{\mathrm{flat},A}(\mathbf{k}_{\perp})$ belongs to $(d-1)$-dimensional class $(s+2)$.
Therefore, $K_{\mathrm{FS}}^{\mathrm{AZ}+I}(s+2,d-1)$ can be considered as a classification of gapless single-particle entanglement spectrum for the $d$-dimensional bulk insulator of class $s$.
Then, the above relation means that the classification of the bulk gapped topological phases and the classification of the gapless single particle entanglement spectra are described by the same $K$ group. We can thus say that there is a ``bulk-edge correspondence" between the bulk Hamiltonian and the entanglement spectrum for the real AZ$+I$ symmetry classes; we call this variant of the bulk-edge correspondence as {\it bulk-entanglement spectrum correspondence }.

In the next section, we explicitly construct models and calculate their entanglement spectrum for spatial dimensions up to three to demonstrate this bulk-entanglement spectrum correspondence.

\section{Models}
\label{sec:models}

We now construct explicit examples of non-trivial topological phases within the AZ$+I$ classification and calculate their entanglement spectra. We first note that, when the $K$-group is $\mathbb{Z}$ for some symmetry class, the topology of these phases are described by the same topological invariants as in class A or AIII, which do not involve $PT$ or $PC$ symmetry. This implies that, even in AZ$+I$ classification, if the topology is classified by $\mathbb{Z}$ or $2\mathbb{Z}$, the ordinary edge spectrum is gapless which holds from the conventional bulk-edge correspondence. When the ordinary edge spectrum is gapless, the edge spectrum of the flattened Hamiltonian, which is nothing but the entanglement spectrum, is also gapless and thus the bulk-entanglement spectrum correspondence holds. 

The difference between AZ and AZ$+I$ classification is thus manifest in symmetry classes where the topology is $\mathbb{Z}_2$. We will now explicitly construct topologically nontrivial models within the AZ$+I$ classification in each dimension where the bulk topology is $\mathbb{Z}_2$.

\subsection{One-dimensional class AI$^{\prime}$ system}

One-dimensional class AI$^\prime$ topological insulators are classified by the $\mathbb{Z}_2$-valued topological number. We denote the $PT$ symmetry as $PT=U_TK$, where $U_T$ is a unitary matrix and $K$ is complex conjugation. If $U_T$ does not depend on $k$, the Berry phase along the 1D Brillouin zone is quantized, and serves as the topological invariant. If $U_T$ depends on $k$, quantized topological invariant can be obtained by adding a correction term to the Berry phase (see Appendix \ref{AI_Berry} for details).
Here, we consider the following model with the quantized Berry phase:
\begin{align}
\mathcal{H}_{\textrm{1D-AI}^{\prime}}
=&(t_0+t_1\cos k)\sigma_x + t_1\sin k \sigma_y
\notag \\
&\ +t_2\cos 2k \mathbb{I} . 
\label{1D_AI_X=0-1}
\end{align}
Here, $\sigma_i(i=x,y,z)$ are Pauli matrices, and $PT$ symmetry is represented by $PT=\sigma_x K$. 
This model is a modification of the well known Su-Schrieffer-Heeger (SSH) model\cite{PhysRevLett.42.1698}, achieved by adding a term proportional to the identity matrix to distort the energy bands.

\begin{figure}[t]
\centerline{\includegraphics[width=8.5cm,clip]{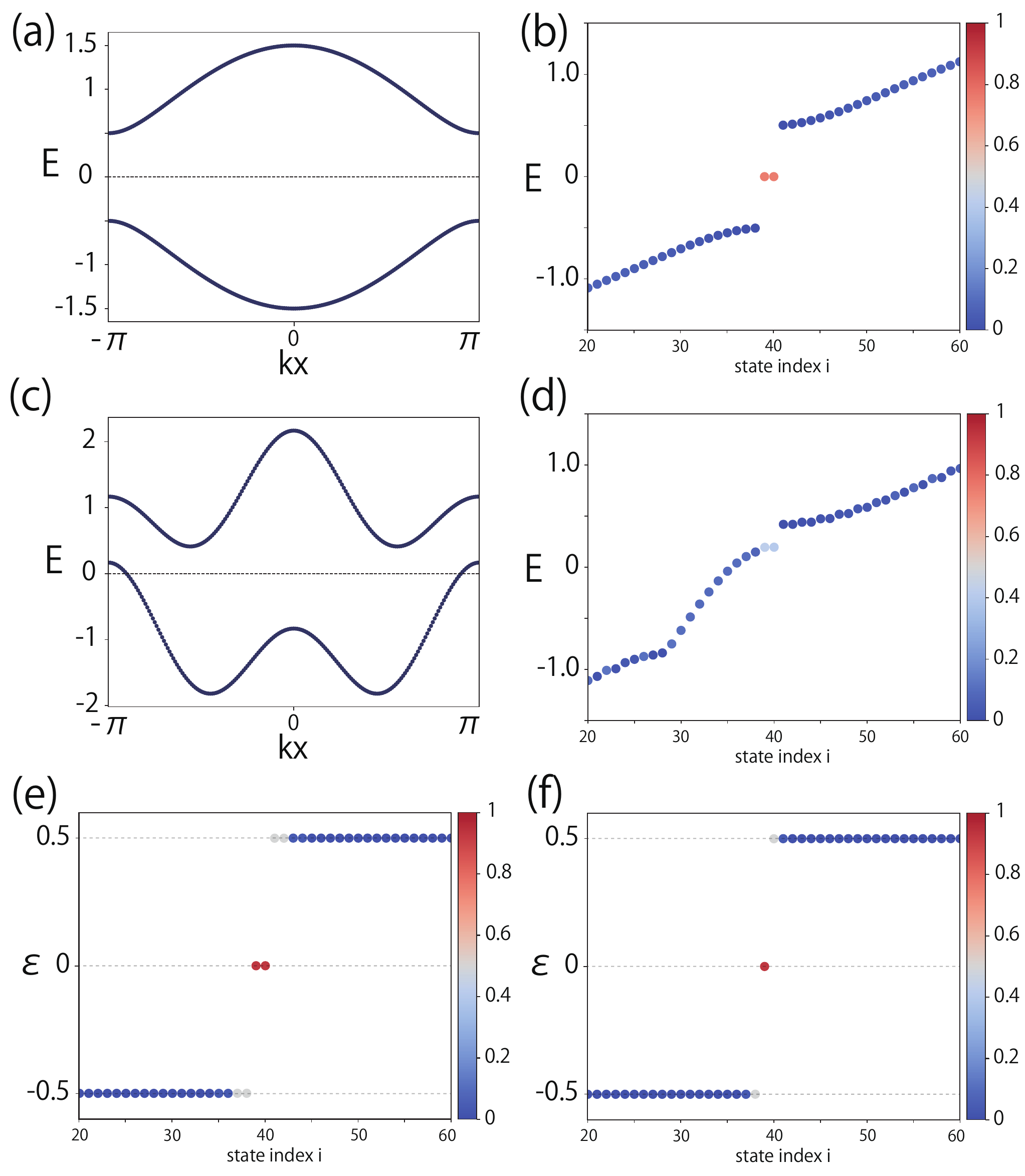}}
  \caption{
  Numerical calculations for the one-dimensional class AI$^\prime$ model [see Eq.~(\ref{1D_AI_X=0-1})]. (a)-(b) The bulk energy band of the standard SSH model [(a)] and the energy spectrum under open boundary conditions [(b)] with parameters $(t_0,t_1,t_2) = ( \frac{1}{2}, 1, 0)$. (c)-(f) The bulk energy band [(c)], energy spectrum under open boundary condition[(d)], and entanglement spectrum [(e,f)] with parameters $(t_0,t_1,t_2) = ( \frac{1}{2}, 1, \frac{2}{3})$. In (e) and (f), the case using the entanglement cut of Fig.~\ref{entanglement_cut}(a) with PBC and Fig.~\ref{entanglement_cut}(b) with OBC are shown, respectively. Colors in (b), (d), (e) and (f) represent the squared amplitudes of the states in the unit cells [(b), (d)] at both ends, or [(e), (f)] at the ends of the entanglement cuts.
  }
    \label{fig:1d_AI_X=1}
\end{figure}

When $t_2=0$, this model is identical to the SSH model, which belongs to class BDI in AZ classification. The bulk energy spectrum is shown in Fig.~\ref{fig:1d_AI_X=1}(a). Upon calculating the energy spectrum under the open boundary conditions, zero energy edge modes appear in the energy gap [see Fig.~\ref{fig:1d_AI_X=1}(b)].

When $t_2$ becomes non-zero, the model is in the class AI in the AZ classification, which is trivial. However, in AZ$+I$ classification, the model is in the class AI$^\prime$, and can have $\mathbb{Z}_2$ topology. 
The energy band is distorted as shown in Fig.~\ref{fig:1d_AI_X=1}(c). Generally, edge states present when $t_2 = 0$ move away from the zero energy and hybridize with the bulk band [Fig.~\ref{fig:1d_AI_X=1}(d)]. 
Therefore, conventional edge spectra may not have edge modes. On the other hand, the entanglement spectrum has a zero energy edge mode if the Berry phase plus a correction term is $\pi$ [see Fig.~\ref{fig:1d_AI_X=1}(e), (f)]. For reference, Fig.~\ref{fig:1d_AI_X=1}(e) and Fig.~\ref{fig:1d_AI_X=1}(f) show calculations for scenarios using the entanglement cut illustrated in Fig.~\ref{entanglement_cut}(a) under periodic boundary conditions (PBC) and in Fig.~\ref{entanglement_cut}(b) under open boundary conditions (OBC), respectively. In Fig.~\ref{fig:1d_AI_X=1}(e), which corresponds to two entanglement cuts, two zero-energy edge modes are observed. In contrast, Fig.~\ref{fig:1d_AI_X=1}(f), which corresponds to a single entanglement cut, exhibits only one zero-energy edge mode. These edge modes are fixed at $E=0$ thanks to $PT\Xi$, which serves as an effective $PC$ symmetry of the entanglement spectrum.

\subsection{One-dimensional class BDI$^{\prime}$ system}\label{Sec:Model_1D_BDI}

Topological phases in class BDI$^{\prime}$ are not classified by the Berry phase or the conventional winding number. As detailed in Ref.~[\onlinecite{PhysRevB.96.155105}], the topological number is defined based on the phase winding of the eigenvalues of an orthogonal matrix, $q(k)$, which is derived from the off-diagonal blocks of the flattened Hamiltonian. 

\begin{figure}[t]
\centerline{\includegraphics[width=8.5cm,clip]{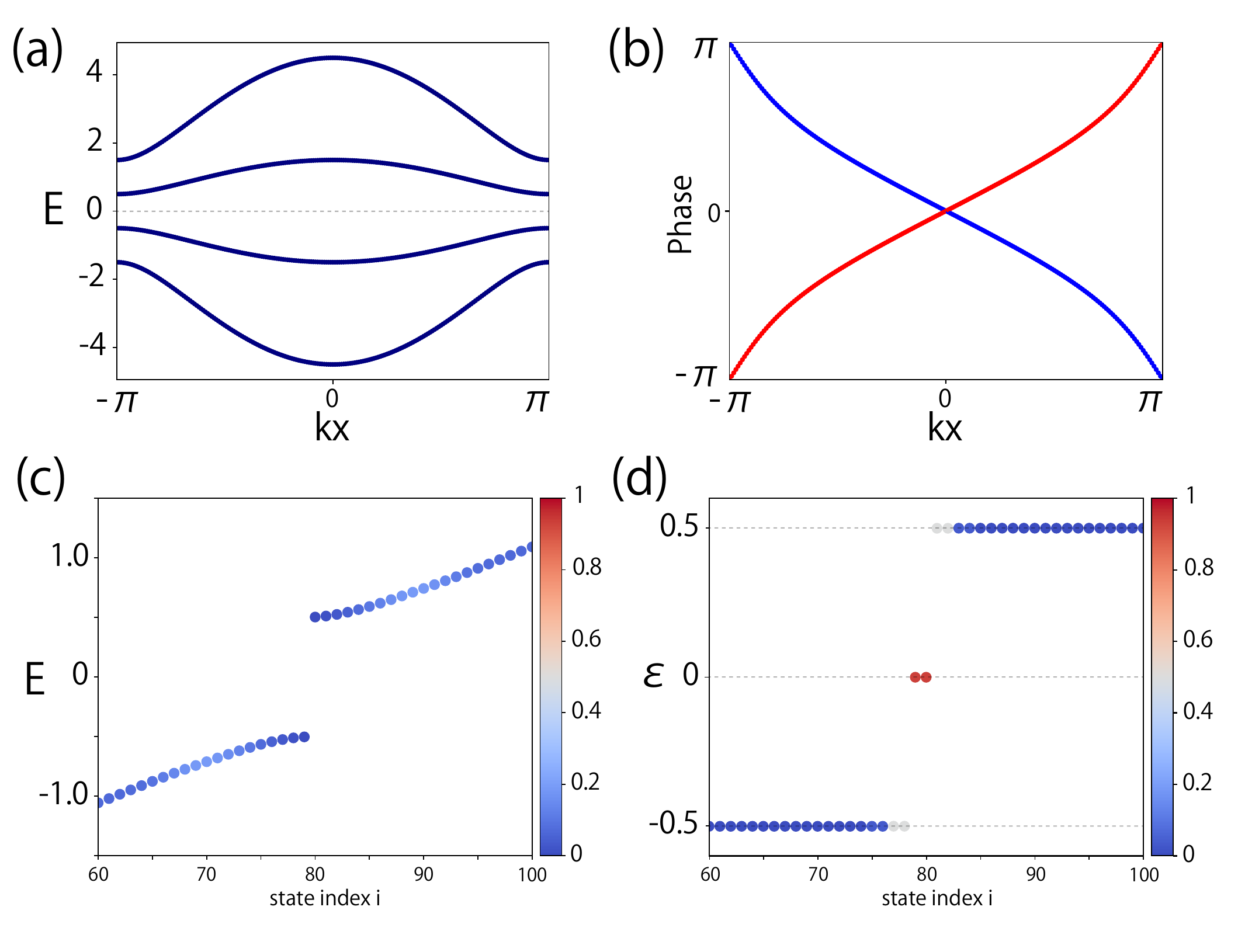}}
  \caption{Numerical calculations for the one-dimensional BDI model [Eq.~(\ref{1D_BDI_X=1})]. (a) Bulk band structure. (b) Phase winding of the eigenvalues of $q(k)$. There are two eigenvalues of $q(k)$; the phase of the eigenvalue of $q(k)$ which increases as $k$ increases is depicted in red, whereas the phase which decreases as $k$ increases is depicted in blue. (c) Edge energy spectrum. (d) entanglement spectrum. 
  Colors in (c) and (d) represent the squared amplitudes of the states in the unit cells (c) at both ends, or (d) at the end of the entanglement cut.
  }
    \label{fig:1d_BDI_X=1}
\end{figure}

We use the following one-dimensional class BDI$^{\prime}$ model: 
\begin{align}
\mathcal{H}_{\textrm{1D-BDI}^{\prime}}
=&\begin{pmatrix}
\bm{0} & Q(k) \\
Q(k)^{T} & \bm{0}
\end{pmatrix}
\\
=&(1+2\cos k)\tau_x + 2\sin k\ \tau_y\sigma_y\notag \\
&-\frac{2+\cos k}{2}\tau_x\sigma_z-\frac{\sin k}{2}\tau_x\sigma_x,
\label{1D_BDI_X=1}
\end{align}
where
\begin{align}
    &Q(k)=
    \\
    &(1+2\cos k)\mathbb{I}+2\sin k(-i\sigma_y)-\frac{2+\cos k}{2}\sigma_z-\frac{\sin k}{2}\sigma_x. \notag 
\end{align}
This model has $PT=K$, and the chiral symmetry $\Gamma=\tau_z$. 
The band structure is shown in Fig.~\ref{fig:1d_BDI_X=1}(a). 
The topological invariant can be obtained by the following procedure. First, we flatten the Hamiltonian $\mathcal{H}_{\textrm{1D-BDI}^{\prime}}$, which should keep the block anti-diagonal form. Calling the upper right block of the flattened Hamiltonian as $q(k)$, the phases of the eigenvalues of $q(k)$ show winding of 1 as one changes $k$ from $-\pi$ to $\pi$, as shown in Fig.~\ref{fig:1d_BDI_X=1}(b)~\footnote{
In our case $\det q(k) = +1$. However, when $\det q(k) = -1$, it is the phase winding of the eigenvalue of $q(k)\cdot\text{diag}[1,\cdots,1,-1]$, rather than that of $q(k)$, that exhibits nontrivial winding and is utilized as a topological invariant.}.
Here, the eigenvalues of $q(k)$ appear in pairs of $e^{\pm i\theta(k)}$, where $e^{+i\theta(k)}$ and $e^{-i\theta(k)}$ have windings in opposite directions. We count the winding number of the one with the positive winding (depicted in red in Fig.~\ref{fig:1d_BDI_X=1}(b)).
The energy spectrum under open boundary conditions generally has an energy gap as shown in Fig.~\ref{fig:1d_BDI_X=1}(c). 
On the other hand, as shown in Fig.~\ref{fig:1d_BDI_X=1}(d), the entanglement spectrum has zero-energy edge modes. Note that the degeneracy of edge modes is a Kramers degeneracy due to the effective time-reversal symmetry $PC\Xi$, which squares to $-\mathbb{I}$.

\subsection{2D class CI$^{\prime}$ and AI$^{\prime}$ system}

\begin{figure}[t]
\centerline{\includegraphics[width=8.5cm,clip]{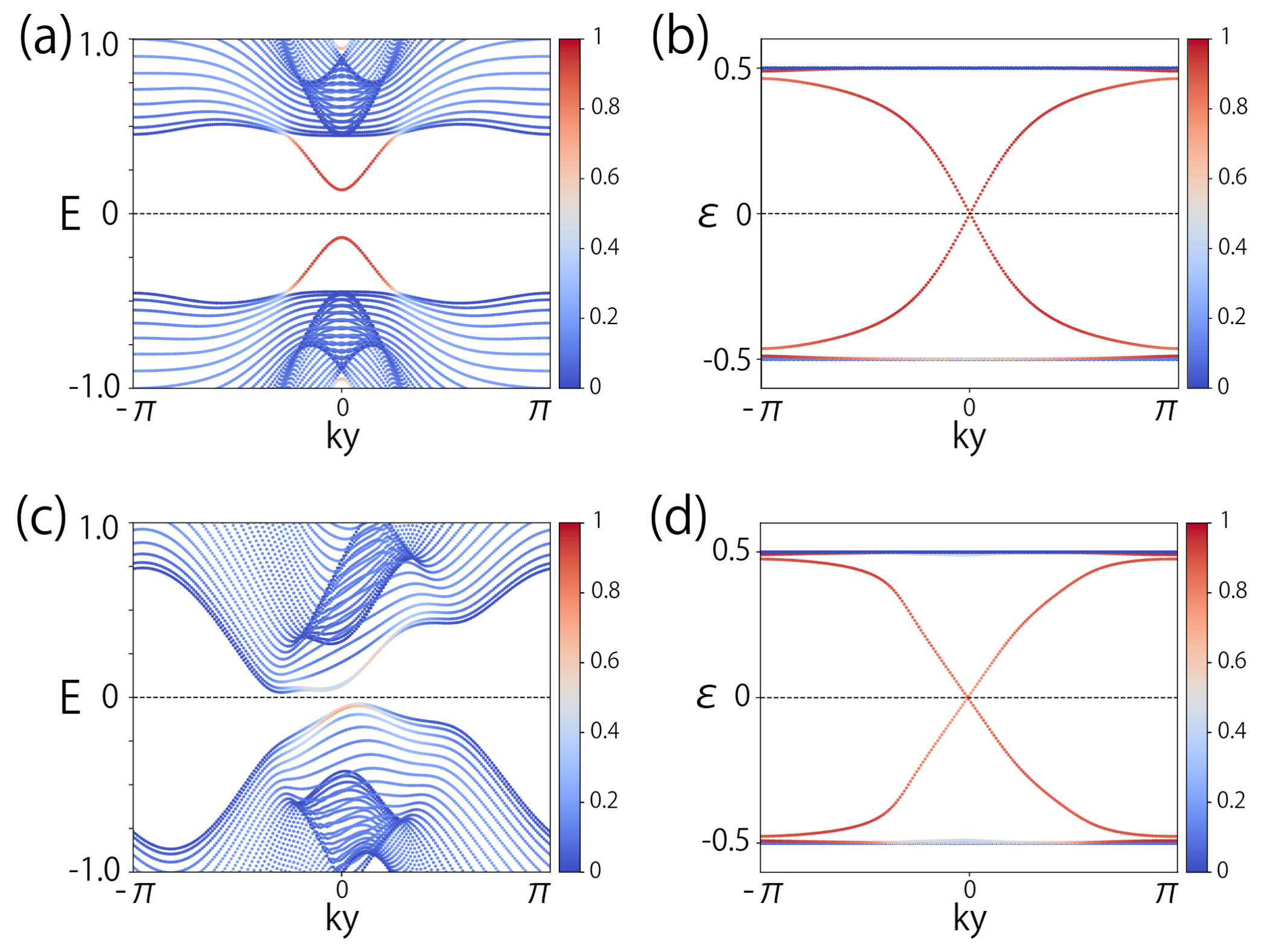}}
  \caption{Numerical calculations for the two-dimensional (a,b) CI$^{\prime}$ model [Eq.~(\ref{2D_CI_X=1})], and (c,d) AI$^{\prime}$ model [Eq.~(\ref{2D_AI_X=1})]. (a,c) Energy spectrum with open boundary condition. (b,d) entanglement spectrum. Colors represent the squared amplitudes of the states in the unit cells (a), (c) at both ends, or (b), (d) at the end of the entanglement cut.
  }
    \label{fig:2d_CI-AI_X=1}
\end{figure}

Two-dimensional class CI$^{\prime}$ systems can be considered as a special case of class AI$^{\prime}$ systems, and thus, we will discuss these two together in this subsection. 
First, we start from the following two-dimensional class CI$^{\prime}$ model: 
\begin{align}
\mathcal{H}_{\text{2D-CI}^{\prime}}
=&\sin k_x\sigma_x+\sin k_y\sigma_y\tau_y
\notag \\
&\ +(1-\cos k_x-\cos k_y)\sigma_z + \frac{\sigma_z\tau_z}{2}+\frac{\tau_x}{4}
\label{2D_CI_X=1}.
\end{align}
This Hamiltonian possesses $PT=K$, and was proposed as a model for Stiefel-Whitney insulator\cite{PhysRevLett.121.106403,PhysRevB.108.075129}. 
This Hamiltonian additionally has chiral symmetry $\Gamma=\sigma_y\tau_z$, and belongs to class CI$^{\prime}$. 
By adding a perturbation that breaks chiral symmetry $\Gamma$ while maintaining $PT$ symmetry, the class of the system can be changed from CI$^{\prime}$ to AI$^{\prime}$:
\begin{align}
\mathcal{H}_{\text{2D-AI}^{\prime}}
=\mathcal{H}_{\text{2D-CI}^{\prime}} +\frac{\sigma_x\tau_z}{2}+\frac{\sigma_z}{5}+\frac{\sigma_x\tau_x}{5}
\label{2D_AI_X=1}.
\end{align}
As shown in Figs.~\ref{fig:2d_CI-AI_X=1}(a) and (c), when calculating the energy spectra under open boundary conditions for these models, there is an energy gap in both cases.
On the other hand, both entanglement spectra exhibit gapless linear dispersion, as shown in Figs.~\ref{fig:2d_CI-AI_X=1}(b) and (d). 
The stability of these gapless points is guaranteed by the sign change of the Pfaffian across the gapless points, which is discussed in detail in our previous work [\onlinecite{PhysRevB.108.075129}].

\subsection{3D class C$^{\prime}$ system}

\begin{figure}[t]
\centerline{\includegraphics[width=8.5cm,clip]{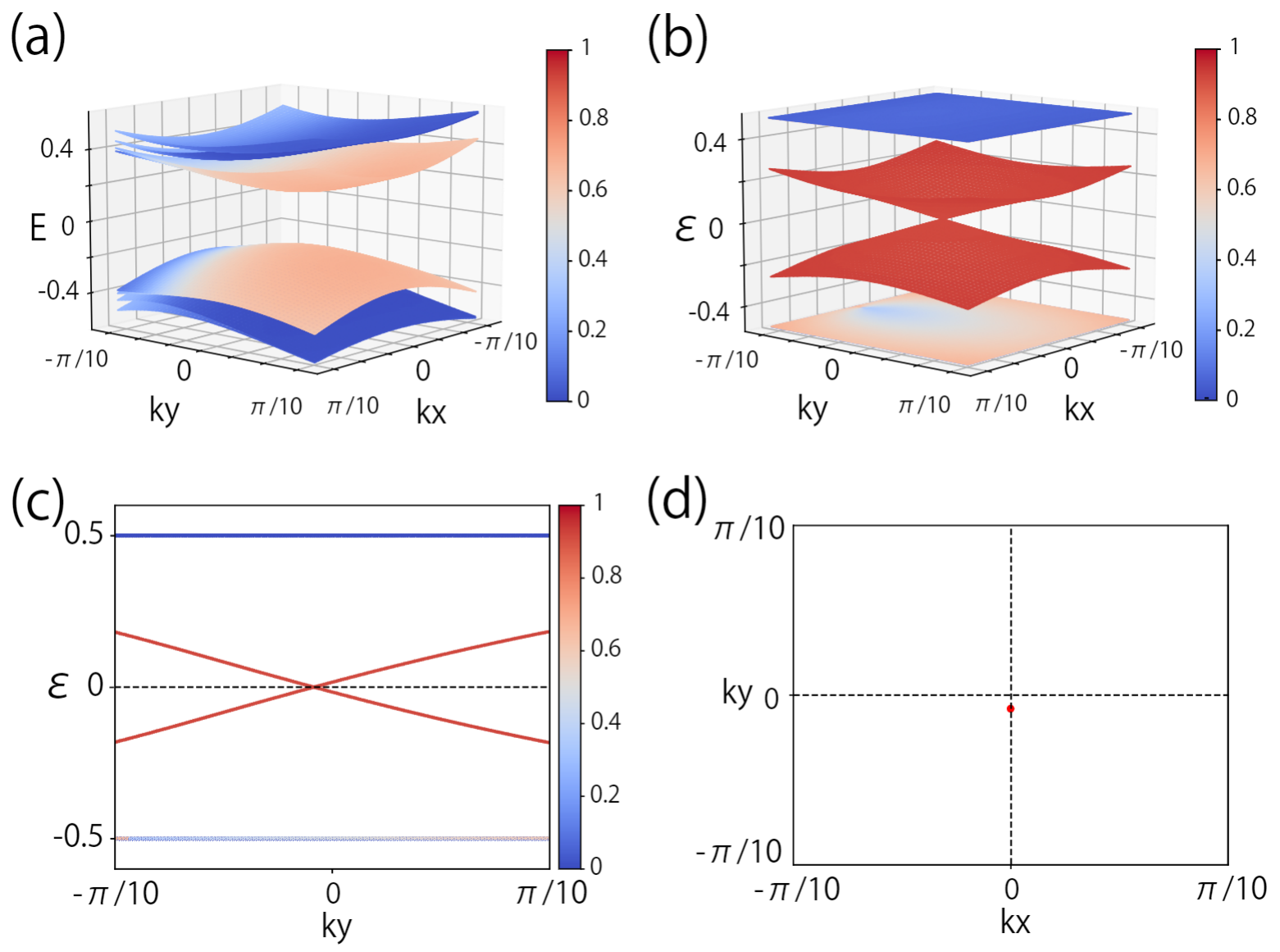}}
  \caption{Numerical calculations for the three-dimensional C$^{\prime}$ model [Eq.~(\ref{3D_C_X=1})]. 
  (a) Energy spectrum under open boundary conditions, with 16 unit cells in the $z$-direction ($N_z$=16). (b) entanglement spectrum. (c) entanglement spectrum along $k_x=0$. (d) Position of the 2D Weyl point in the entanglement spectrum. The 2D Weyl point is located at $(k_x,k_y)\approx(0,-0.025)$. Colors in (a), (b) and (c) represent the squared amplitudes of the states in the unit cells (a) at both ends, or (b),(c) at the end of the entanglement cut.
  }
    \label{fig:3d_C_X=1}
\end{figure}

We use the following three-dimensional four-band model: 
\begin{align}
\mathcal{H}_{\text{3D-C}^{\prime}}
=&-(\sin k_x\sigma_x\tau_x + \sin k_y\sigma_y\tau_x+\sin k_z\sigma_z\tau_x)
\notag \\
&\ +(2.5-\cos k_x -\cos k_y -\cos k_z)\tau_y
\notag \\
&\ +\frac{\sigma_x\tau_z}{5}+\frac{\sigma_z}{5}
\label{3D_C_X=1}.
\end{align}
This model belongs to class C$^{\prime}$ and has $PC=i\sigma_y K$. 
As shown in Fig.~\ref{fig:3d_C_X=1}(a), the edge spectrum is gapped. On the other hand, as shown in Figs.~\ref{fig:3d_C_X=1}(b-d), the entanglement spectrum has a gapless linear dispersion consisting of two bands, called 2D Weyl point. 
From the Table\ref{tab:Bulk_edge_AZ+I}, the entanglement spectrum belongs to class AI$^{\prime}$, and the 2D Weyl point is protected by quantized "Berry phase" which is modified by the correction term discussed in Appendix \ref{3D_C_Berry}.

\subsection{3D class CI$^{\prime}$ system}

\begin{figure}[t]
\centerline{\includegraphics[width=8.5cm,clip]{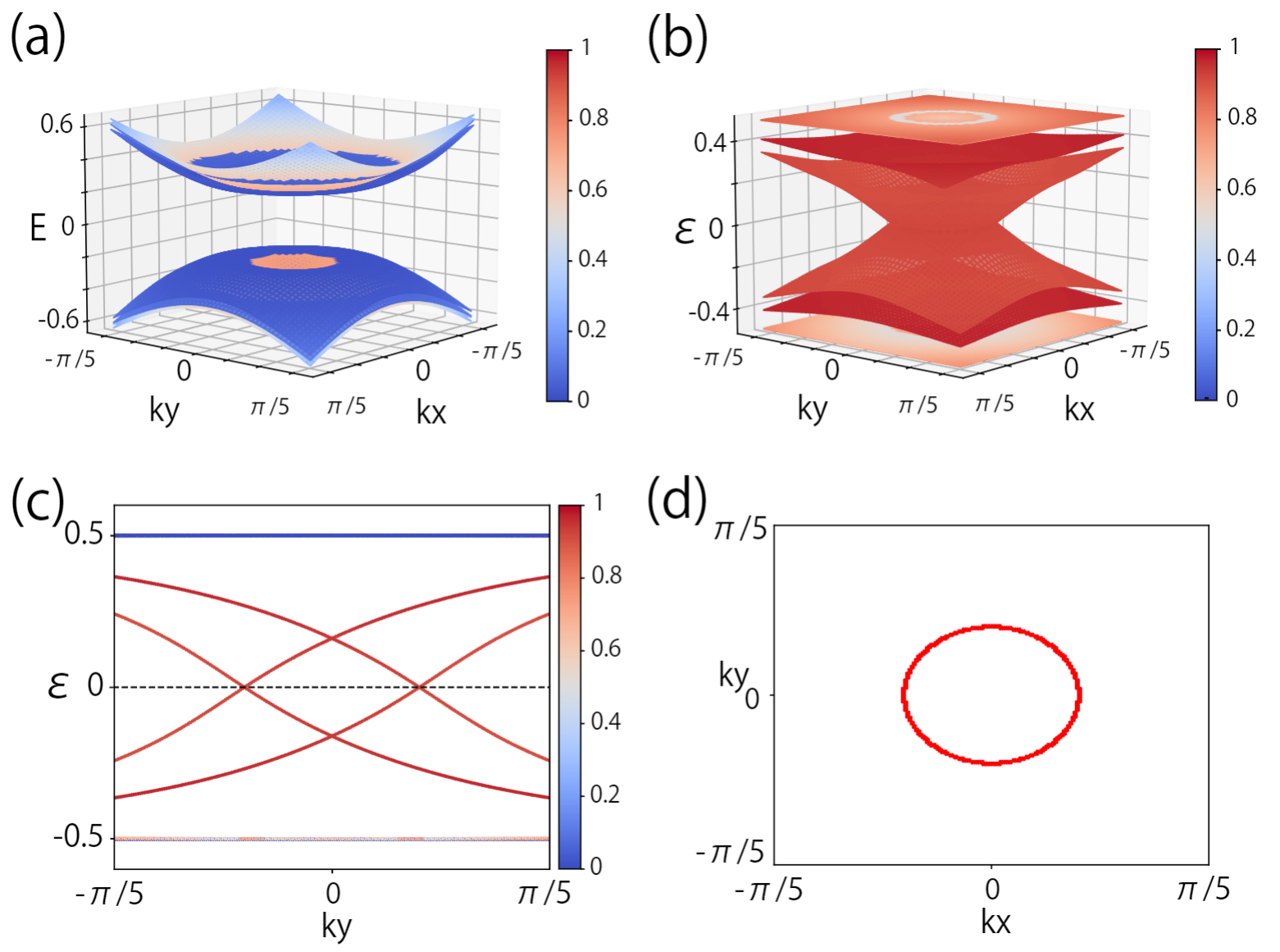}}
  \caption{Numerical calculations for the three-dimensional CI$^{\prime}$ model [Eq.~(\ref{3D_CI_X=1})]. 
  (a) Energy spectrum under open boundary conditions, with 16 unit cells in the $z$-direction ($N_z$=16). (b) entanglement spectrum. (c) entanglement spectrum along $k_x=0$. (d) Position of the nodal line in the entanglement spectrum. Colors in (a), (b) and (c) represent the squared amplitudes of the states in the unit cells (a) at both ends, or (b),(c) at the end of the entanglement cut.
  }
    \label{fig:3d_CI_X=1}
\end{figure}

We use the following three-dimensional eight-band model: 
\begin{align}
\mathcal{H}_{\text{3D-CI}^{\prime}}
=&-(\sin k_x\sigma_x\tau_x\mu_y + \sin k_y\sigma_z\tau_x\mu_y+\sin k_z\mu_x)
\notag \\
&\ +(2.5-\cos k_x -\cos k_y -\cos k_z)\tau_z\mu_y
\notag \\
&\ +\frac{\tau_x\mu_y}{4}-\frac{\tau_x\mu_x}{4}. 
\label{3D_CI_X=1}
\end{align}
Here, $\mu_i(i=1,2,3)$ are Pauli matrices.
This model belongs to class CI$^{\prime}$, and has $PT=\mu_xK$ and $\Gamma=\mu_z$.  
As shown in Fig.~\ref{fig:3d_CI_X=1}(a), the edge spectrum is gapped. On the other hand, the entanglement spectrum has a gapless 2D nodal-line boundary state as shown in Figs.~\ref{fig:3d_CI_X=1}(b-d). 
From the Table\ref{tab:Bulk_edge_AZ+I}, the entanglement spectrum belongs to class BDI$^{\prime}$, and thus the nodal-line state is protected by two $\mathbb{Z}_2$ invariants, 0-dimensional Pfaffian invariant and 1-dimensional phase-winding invariant. The Pfaffian takes different signs, $\pm1$, inside and outside the nodal line, and the robustness of the degeneracy is guaranteed by the non-trivial $\mathbb{Z}_2$ topological invariant defined for a one-dimensional loop surrounding the nodal line.

\section{Fragile topology}\label{sec:fragile}
In this section, we discuss the possibility that bulk-entanglement spectrum correspondence also holds for so-called fragile topological insulators. 
Fragile topological insulators are topological insulators whose classification also depends on the number of occupied bands\cite{PhysRevLett.121.126402}. 
(We note that if the classification of fragile topological insulators also depends on the number of unoccupied bands\cite{PhysRevLett.126.216404}, they are called delicate topological insulators.)
In two-dimensional class AI$^{\prime}$ systems, there exists a fragile topological insulator called Euler insulator\cite{PhysRevB.96.155105,PhysRevX.9.021013,PhysRevLett.125.053601,bouhon2020non,PhysRevB.103.205303,zhao2022quantum}, which is characterized by an integer called the Euler number. As we numerically found in our previous paper~\cite{PhysRevB.108.075129}, the Euler insulator with the Euler number $n$ has $n$ modes crossing the zero energy in the entanglement spectrum, including the multiplicity at the zero energy. In conventional topological insulators characterized by the Chern number, one does not need to count the multiplicity to find the edge topological invariant, and thus the Euler insulator shows the bulk-edge (or bulk-entanglement spectrum) correspondence which is qualitatively different from the conventional topological insulators. 

In the AZ$+I$ classification, for one- and two-dimensional systems, previous research [\onlinecite{PhysRevB.96.155105}] has shown the classification of fragile topological phases using homotopy groups. 
They have found that, in classes AI$^{\prime}$ and BDI$^{\prime}$ in one dimension and classes AI$^{\prime}$ and CI$^{\prime}$ in two dimensions, the bulk topological invariant changes from $\mathbb{Z}_2$ to $\mathbb{Z}$ if the number of occupied bands takes specific values.

Class AI$^{\prime}$ in two dimensions is the case which was treated in our previous work~\cite{PhysRevB.108.075129}. Such systems are fragile topological insulators when the number of occupied bands is two and is described by the Euler number. In this previous work, we have also confirmed fragile nature of the entanglement spectrum, namely the zero energy crossing of the entanglement spectrum opens a gap once trivial bands are added and mixed with the occupied bands.

The situation of class CI$^\prime$ in two dimensions with two occupied band is essentially the same as class AI$^\prime$ because they are both described by the Euler number.

Next we discuss one dimensional fragile and delicate topological insulators in AZ$+I$ classification, which are classes AI$^{\prime}$ and BDI$^{\prime}$.

\subsection{1D class AI$^{\prime}$ system}

Fragile topological insulators in one-dimensional class AI$^{\prime}$ systems appear when there is exactly one occupied band and one unoccupied band. In this scenario, the Hamiltonian is described by three Pauli matrices, along with an identity matrix. More precisely, due to the $PT$ symmetry constraint, the total degrees of freedom for the Pauli matrices are limited to two. For example, if $PT = \sigma_x K$, $\sigma_z$ is prohibited, allowing only the two Pauli matrices, $\sigma_x$ and $\sigma_y$, to be used. Subsequently, one can define a two-dimensional vector using coefficients $a_x$ and $a_y$ of $\sigma_x$ and $\sigma_y$, respectively. This two-dimensional vector enables the determination of a $\mathbb{Z}$-valued topological number from the number of times the vector encircles the origin.

\begin{figure}[t]
\centerline{\includegraphics[width=9cm,clip]{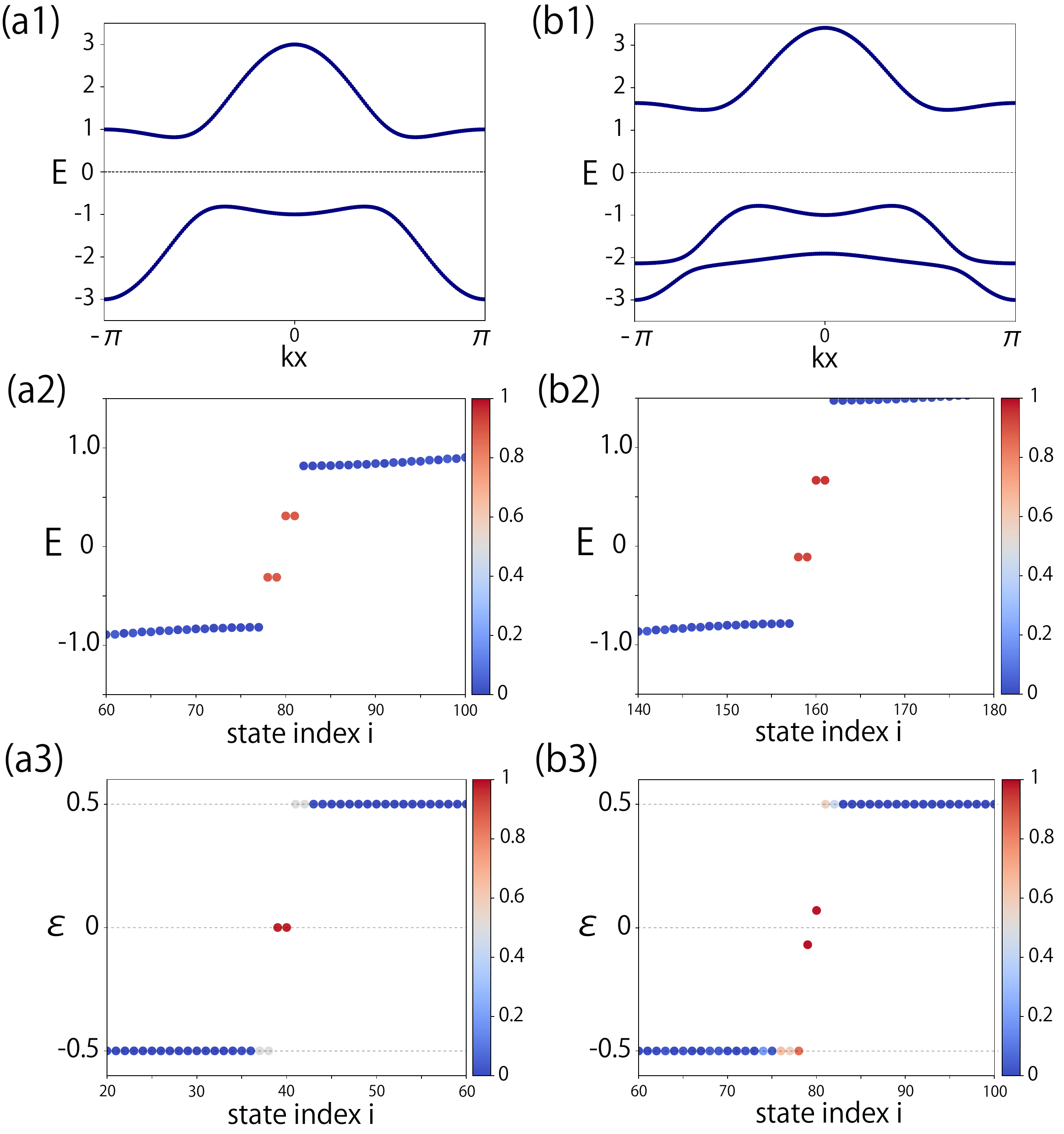}}
  \caption{Numerical calculations for the one-dimensional fragile topological insulator model with class AI$^{\prime}$ symmetry in (a) Eq.~(\ref{1D_AI_X=2}) and in (b) Eq.~(\ref{1D_AI_fragile_X=2}). (a1, b1) Bulk band structure. (a2, b2) Edge energy spectrum. (a3, b3) entanglement spectrum. The colors in (a2) and (b2) [(a3) and (b3)] represent the sum of the squared amplitudes of the wave function in two unit cells from both ends [from the entanglement cut]. 
  }
    \label{fig:1d_AI_X=2}
\end{figure}

The following two-band model is an example of such fragile topological insulators:
\begin{align}
\mathcal{H}_{\textrm{1D-AI}^{\prime}, {F}}
&=\frac{1+3\cos2k}{2}\sigma_x+\frac{3\sin2k}{2}\sigma_y+\cos k\mathbb{I}.
\label{1D_AI_X=2}
\end{align}
This model has $PT=\sigma_x K$, and the winding number is two.
The band structure is shown in Fig.~\ref{fig:1d_AI_X=2}(a1). The edge spectrum generally has no zero energy state as shown in Fig.~\ref{fig:1d_AI_X=2}(a2), but there are two zero-energy entanglement edge modes as shown in Fig.~\ref{fig:1d_AI_X=2}(a3). We note that, from the stable bulk-entanglement spectrum correspondence we have established in this paper, one-dimensional class AI$^{\prime}$ allows only $\mathbb{Z}_2$ topological invariant and thus two zero modes should be unstable against general perturbations.

The appearance of two zero modes here can be understood in the following way. The last term in the Hamiltonian in Eq.~(\ref{1D_AI_X=2}) can be removed without changing the eigenstates and the flattened Hamiltonian. The Hamiltonian with the last term removed has $\sigma_z$ as a chiral symmetry, and thus belongs to class CI$^{\prime}$ where $\mathbb{Z}$-topological invariant is allowed. The winding number of the vector $(a_x,a_y)$ is exactly this $\mathbb{Z}$-topological invariant, and from the stable bulk-entanglement spectrum correspondence of class CI$^{\prime}$, the number of zero-energy entanglement spectrum edge modes should be equal to the winding number. The fragile bulk-entanglement spectrum correspondence for class AI$^{\prime}$ systems can thus be understood through the stable bulk-entanglement spectrum correspondence for class CI$^{\prime}$ systems.

Contrary to the zero-energy edge modes found in the entanglement spectrum of stable topological insulators, those in fragile topological insulators become unstable with changes in the number of occupied bands. 
To illustrate this, we consider the following model:
\begin{align}
\mathcal{H}_{\textrm{1D-AI}^{\prime}, {F^{\prime}}}
&=\begin{pmatrix}
\mathcal{H}_{\textrm{1D-AI}^{\prime}, {F}} & \Delta_{\text{AI}^{\prime}} \\
\Delta_{\text{AI}^{\prime}}^{\dagger} & -1.5
\end{pmatrix},
\label{1D_AI_fragile_X=2}
\\
\Delta^{\dagger}_{\text{AI}^{\prime}}&=\begin{pmatrix}
e^{ik}, & e^{-ik}
\end{pmatrix}.
\end{align}
This model has $PT=\bigl{(}\begin{smallmatrix}
\sigma_x & \\
 & 1
\end{smallmatrix}\bigr{)}K$. 
The band structure is shown in Fig.~\ref{fig:1d_AI_X=2}(b1).
Since adding a trivial band changes the fragile topological insulator into a trivial insulator, both the edge spectrum and the entanglement spectrum open an energy gap, as shown in Fig.~\ref{fig:1d_AI_X=2}(b2) and (b3).
Thus, both the bulk topology and the entanglement-spectrum are fragile against addition of trivial bands.

\subsection{1D class BDI$^{\prime}$ system}
\begin{figure}[t]
\centerline{\includegraphics[width=8.5cm,clip]{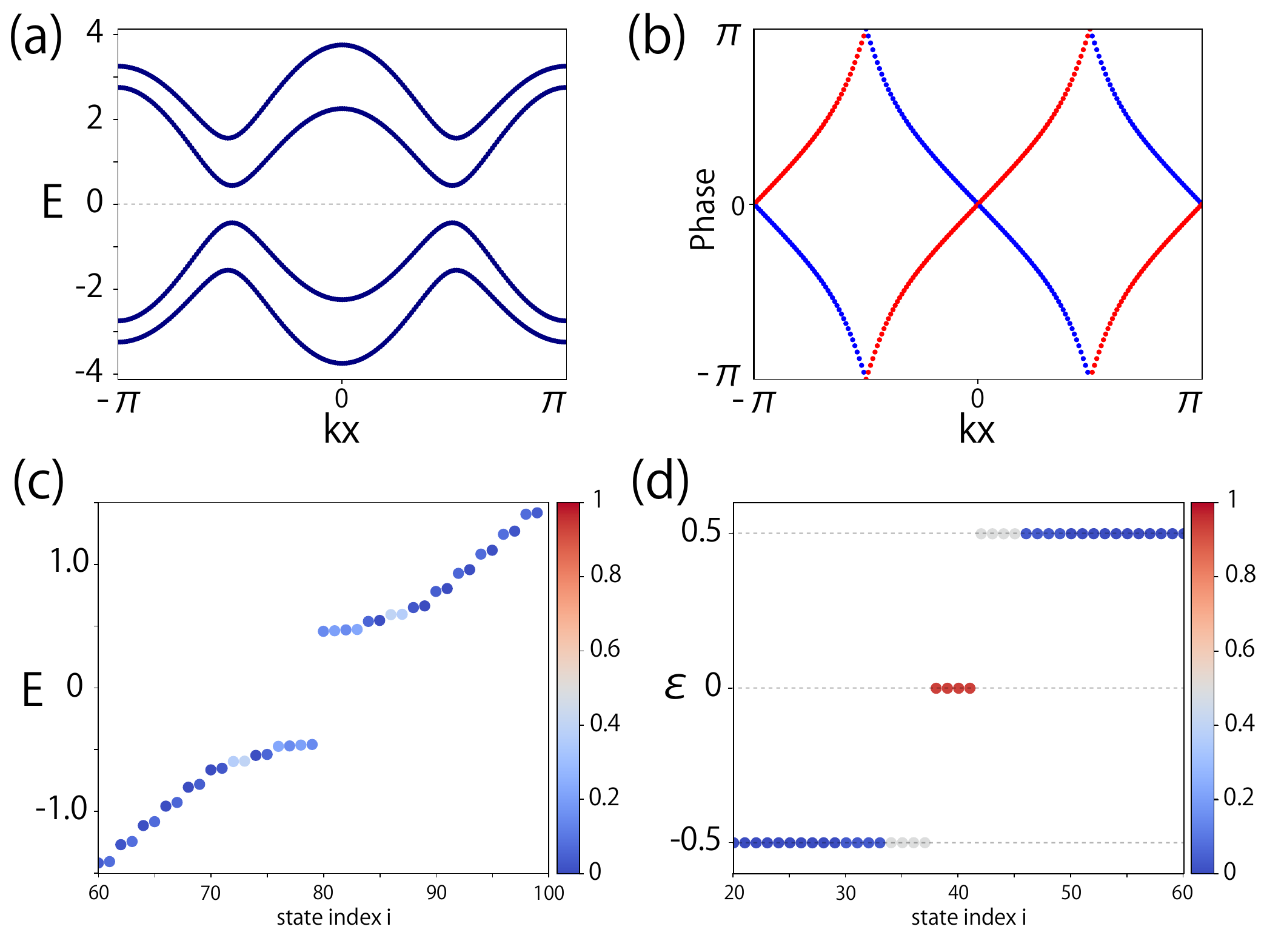}}
  \caption{Numerical calculations for the one-dimensional fragile topological insulator model with class BDI$^{\prime}$ symmetry in Eq.~(\ref{1D_BDI_X=2}). (a) Bulk band structure. (b) Phase winding of the eigenvalues of $q(k)$. The phase of the eigenvalue of $q(k)$ which increases as $k$ increases is depicted in red, whereas the phase which decreases as $k$ increases is depicted in blue. (c) Edge energy spectrum. (d) entanglement spectrum. The colors in (c) [(d)] represent the sum of the squared amplitudes of the wave function in the unit cells at both ends [in two unit cells from the entanglement cut]. 
  }
    \label{fig:1d_BDI_X=2}
\end{figure}

A fragile topological insulator in the one-dimensional class BDI$^{\prime}$ appears when there are exactly two occupied bands. 
The topological number of this insulator is characterized by the winding of the eigenvalues of a certain orthogonal matrix $q(k)$, similar to the case of stable topological insulator\cite{PhysRevB.96.155105}. 
Here, we use the following four-band model:
\begin{align}
&Q_F=(1+2\cos2k)\sigma_0+2\sin2k(-i\sigma_y)\notag \\
&\quad\quad\quad -\frac{2+\cos k}{4}\sigma_z-\frac{\sin k}{4}\sigma_x, 
\\
&\mathcal{H}_{\textrm{1D-BDI}^{\prime}, F}
=\begin{pmatrix}
\bm{0} & Q_F \\
Q_F^{T} & \bm{0}
\end{pmatrix}
\label{1D_BDI_X=2}
\\
&\phantom{\mathcal{H}_{\textrm{1D-BDI}^{\prime}, F}}
=(1+2\cos 2k)\tau_x + 2\sin 2k\ \tau_y\sigma_y\notag \\
&\phantom{\mathcal{H}_{\textrm{1D-BDI}^{\prime}, F}}\quad -\frac{2+\cos k}{4}\tau_x\sigma_z-\frac{\sin k}{4}\tau_x\sigma_x.
\end{align}
Similar to the model in Eq.~(\ref{1D_BDI_X=1}), 
this model also has $PT=K$, and the chiral symmetry $\Gamma=\tau_z$. 
The band structure is shown in Fig.~\ref{fig:1d_BDI_X=2}(a).
The topological invariant can be obtained by flattening the Hamiltonian $\mathcal{H}_{\textrm{1D-BDI}^{\prime}, F}$, in the same manner as described in Sec.\ref{Sec:Model_1D_BDI}. Calling the upper right block of the flattened Hamiltonian as $q(k)$, the phases of the eigenvalues of $q(k)$ show winding of 2 as one changes $k$ from $-\pi$ to $\pi$, as shown in Fig.~\ref{fig:1d_BDI_X=2}(b). 
The edge spectrum generally has no zero energy state as shown in Fig.~\ref{fig:1d_BDI_X=2}(c), but there are four zero-energy entanglement edge modes as shown in Fig.~\ref{fig:1d_BDI_X=2}(d). 

The appearance of four zero-energy entanglement edge modes despite being in class BDI$^{\prime}$ where stable topological invariant is $\mathbb{Z}_2$ can be understood in terms of the stable bulk-entanglement spectrum correspondence of class CI$^{\prime}$ as we describe now. Following the discussion of Ref.~[\onlinecite{PhysRevB.96.155105}], the upper right block $q(k)$ of the flattened Hamiltonian is an element of $O(2)$ and can be written in the form 
\begin{align}
    q(k) = p_1(k) \sigma_0 + p_2 (k) (-i\sigma_y),
\end{align}
where
\begin{align}
    p_1 (k) &= \frac{1+2\cos2k}{\sqrt{(1+2\cos2k)^2 + (2\sin 2k)^2}}, \\
    p_2 (k) &= \frac{2\sin2k}{\sqrt{(1+2\cos2k)^2 + (2\sin 2k)^2}}.
\end{align}
The flattened Hamiltonian itself can therefore be written as
\begin{align}
\begin{pmatrix}
    0 & q(k) \\ q(k)^\dagger & 0
\end{pmatrix}
=
p_1(k) \tau_x + p_2 (k) \tau_y \sigma_y.
\end{align}
This flattened Hamiltonian has an additional unitary symmetry given by $U = \tau_x \sigma_z$.
We can then block-diagonalize the flattened Hamiltonian into sectors with opposite eigenvalues of $U$. 
Concretely, going to a new basis through a unitary matrix
\begin{align}
    M \equiv \frac{1}{\sqrt{2}}\begin{pmatrix}1&0&1&0 \\ 0&1&0&1 \\ 1&0&-1&0\\ 0&-1&0&1\end{pmatrix},
\end{align}
$U$ is diagonalized as $M^{\dagger} U M = \tau_z$ and the flattened Hamiltonian becomes block-diagonalized as
\begin{align}
    & M^{\dagger}\begin{pmatrix}
    0 & q(k) \\ q(k)^\dagger & 0
    \end{pmatrix}M
    =
    \tau_z \left( p_1 (k) \sigma_z + p_2 (k) \sigma_x \right).
\end{align}
The two blocks are opposite, guaranteed by the chiral symmetry of the system, which is $M^{-1}\tau_z M = \tau_x$ in the new basis.
Since the flattened Hamiltonian is block-diagonal and one is given by the opposite of the other, we can analyze the topology of one block, which is $p_1 (k) \sigma_z + p_2 (k) \sigma_x$. This block has an additional chiral symmetry given by $\sigma_y$. Combined with the $PT$ symmetry $K$, we also obtain the particle-hole symmetry $K\sigma_y$ which squares to $-\mathbb{I}$. Thus the block is in class CI$^{\prime}$, which admits $\mathbb{Z}$-topological invariant. Therefore, from the stable bulk-entanglement spectrum correspondence of one-dimensional class CI$^{\prime}$ systems, when the winding of the vector $(p_1 (k), p_2 (k))$ of one block is $N_q$, it is expected that there are $2N_q$ zero-energy entanglement edge modes, counting both blocks. The winding of the vector $(p_1 (k), p_2 (k))$ is nothing but the phase winding of the eigenvalues of $q(k)$, which is the fragile topological invariant of the original Hamiltonian.

\begin{figure}[t]
\centerline{\includegraphics[width=8.5cm,clip]{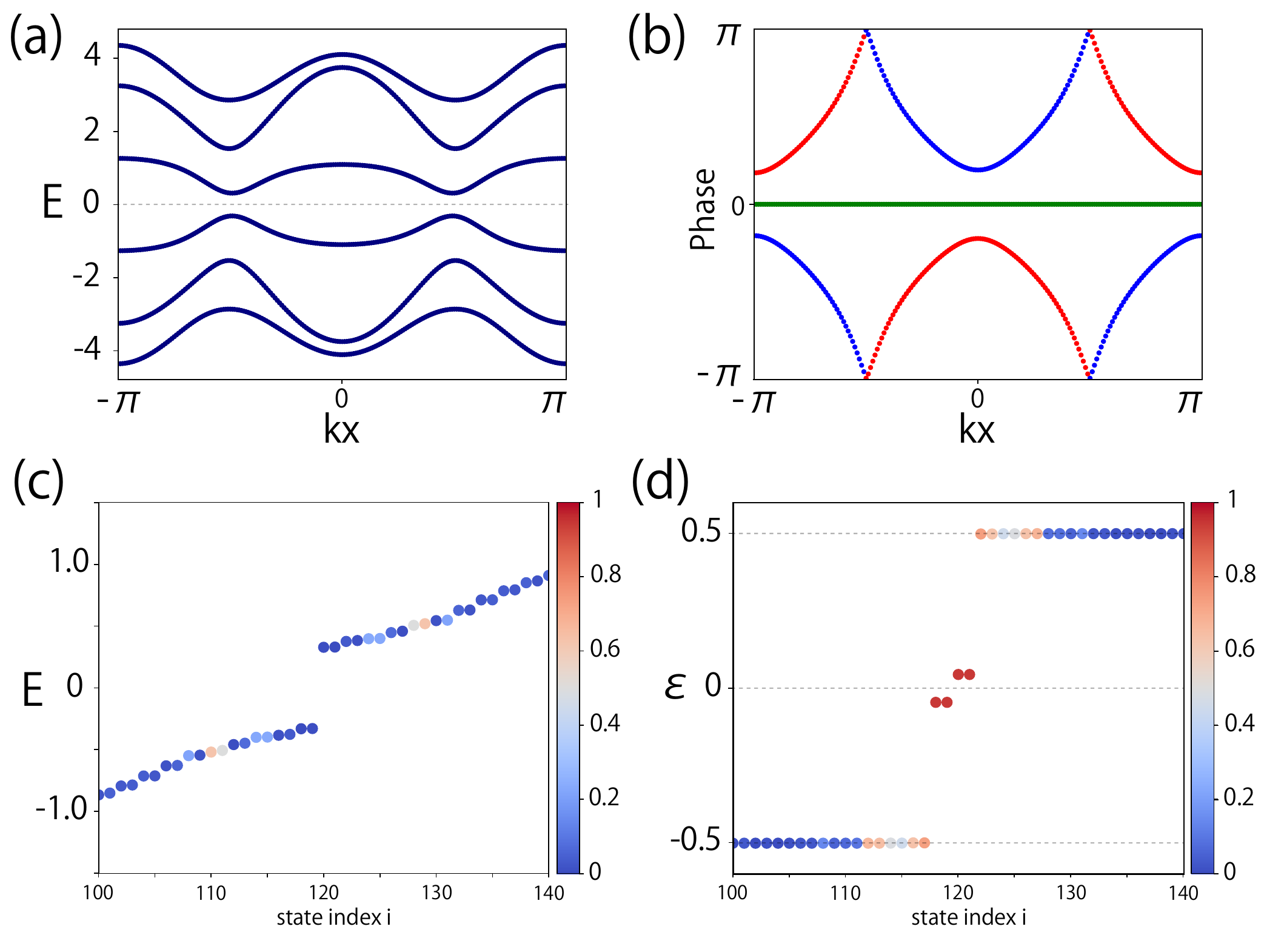}}
  \caption{Numerical calculations for the one-dimensional fragile topological insulator model with class BDI$^{\prime}$ symmetry in Eq.~(\ref{1D_BDI_fragile_X=2}). (a) Bulk band structure. (b) Phase winding of the eigenvalues of $q(k)$. The phase of the eigenvalue of $q(k)$ which increases as $k$ increases is depicted in red, whereas the phase which decreases as $k$ increases is depicted in blue. (c) Edge energy spectrum. (d) entanglement spectrum. The colors in (c) [(d)] represent the sum of the squared amplitudes of the wave function in the unit cells at both ends [in two unit cells from the entanglement cut]. 
  }
    \label{fig:1d_BDI_X=2_fragile}
\end{figure}

To illustrate the instability of the entanglement edge modes, we consider the following model:
\begin{align}
&\mathcal{H}_{\textrm{1D-BDI}^{\prime}, {F^{\prime}}}
=\begin{pmatrix}
\bm{0} & Q_{F^{\prime}} \\
Q_{F^{\prime}}^{T} & \bm{0}
\end{pmatrix},
\label{1D_BDI_fragile_X=2}
\\
&Q_{F^{\prime}}=
\begin{pmatrix}
Q_F & \Delta_{\text{BDI}^{\prime}} \\
\mathbf{0} & 2
\end{pmatrix},\quad \Delta_{\text{BDI}^{\prime}}^{T}=(3\cos k, 2\sin k).
\end{align}
This model is a variant of the one presented in Eq.~(\ref{1D_BDI_X=2}), augmented with a pair of occupied and unoccupied trivial bands. Since the Hamiltonian is a real matrix, this model has $PT=K$ and the chiral symmetry $\Gamma=\text{diag}[\mathbb{I},-\mathbb{I}]$. 
The band structure is shown in Fig.~\ref{fig:1d_BDI_X=2_fragile}(a). The eigenvalues of the upper right block of the flattened Hamiltonian, $q(k)$, shows no phase winding as shown in Fig.~\ref{fig:1d_BDI_X=2_fragile}(b), indicating that the system is trivial. 
Then, both the edge spectrum and the entanglement spectrum have an energy gap as shown in Fig.~\ref{fig:1d_BDI_X=2_fragile}(c) and (d). 


\section{Non-interacting many-body entanglement spectrum}
\label{sec:many-body}
In this section, we discuss the consequence of the bulk-entanglement spectrum correspondence on the many-body entanglement spectrum for systems with non-interacting fermions. 
In the absence of inter-particle interactions, the many-body entanglement spectrum $\{\lambda_{\{n_l\}}\}$ of fermions is uniquely determined from the single-particle entanglement spectrum $\{\epsilon_l\}$ from Eq.~(\ref{MES_ES}) in Sec.~\ref{sec:spes_general}.

Let us start from one dimensional cases. 
If the single-particle entanglement spectrum possesses $N_0$ zero modes, its many-body entanglement spectrum exhibits a $2^{N_0}$-fold degeneracy. This results from the fact that, when $\epsilon_{l_0} = 0$, the values of $\{\lambda_{\{n_l\}}\}$ remain identical whether $n_{l_0}=0$ or $n_{l_0}=1$. 
As demonstrated by several numerical examples, the single-particle entanglement spectra of one-dimensional non-trivial topological insulators and superconductors within the AZ$+I$ classification feature zero energy modes. Consequently, the many-body entanglement spectrum of these insulators exhibits a degeneracy that is a power of two.
The bulk-entanglement spectrum correspondence thus implies that the degeneracy of the many-body entanglement spectrum is topologically protected by symmetries in the AZ$+I$ classes. 

In two or higher dimensions, according to Eq.~(\ref{MES_ES}), if the single-particle entanglement spectrum is gapless, the many-body entanglement spectrum is gapless. Given that the single-particle entanglement spectra of $d$-dimensional topological insulators and superconductors within the AZ$+I$ classification are gapless for $d \geq 2$, it follows that their many-body entanglement spectra are also gapless.
The bulk-entanglement spectrum correspondence thus implies that the many-body entanglement spectrum cannot be gapped out as long as the relevant symmetries in the AZ$+I$ classes are kept.

The above statements on the many-body entanglement are for non-interacting systems. Whether they still hold, or how they should be modified, in the presence of interactions remains a topic for future works.

\section{Conclusion and discussions}
\label{sec:conclusion}
In this paper, we have shown a variant of the bulk-boundary correspondence for $PT$- and $PC$-symmetric topological insulators and superconductors: when the bulk is topologically nontrivial, the entanglement spectrum shows gapless spectrum. We have also constructed concrete models which show this bulk-entanglement spectrum correspondence for all nontrivial symmetry classes in AZ$+I$ classification in dimensions up to three.

The fate of the entanglement spectrum in the presence of inter-particle interactions is an outstanding question. We have shown that, for non-interacting fermions, the many-body entanglement spectrum is gapless as long as the bulk topology is nontrivial in AZ$+I$ classification. If any bulk-entanglement spectrum correspondence remains upon adding inter-particle interaction, i.e., if it is impossible to open a gap in the many-body entanglement spectrum by adding inter-particle interactions which do not close the bulk energy gap, is an interesting question, especially in relation to other works on entanglement spectrum to characterize many-body phases~\cite{PhysRevLett.101.010504,PhysRevB.81.064439,PhysRevLett.104.130502}.

Finally, we point out that the experimental measurement of the entanglement spectrum has recently been reported in IBM quantum computers~\cite{PhysRevLett.121.086808} and ultracold atomic gases~\cite{redon2023realizing}. Reconstruction of the entanglement spectrum of an Euler insulator from its bulk eigenstates has also been realized in trapped ions~\cite{zhao2022quantum}. There are also proposals to measure the entanglement spectrum in various platforms~\cite{PhysRevX.6.041033,dalmonte2018quantum,PhysRevResearch.3.013112}. We expect that study of entanglement spectrum opens a new avenue toward understanding topology and bulk-edge correspondence in systems which were once considered to show no protected edge physics.

\begin{acknowledgments}
We are indebted to Tom\'a\ifmmode \check{s}\else \v{s}\fi{} Bzdu\ifmmode \check{s}\else \v{s}\fi{}ek for showing us how our proof of the bulk-entanglement spectrum correspondence can be extended to the fragile cases given in the paper. This work is supported by JSPS KAKENHI Grant Number JP20H01845, JST PRESTO Grant No. JPMJPR2353, JST CREST Grant Number JPMJCR19T1.
\end{acknowledgments}

\appendix{}

\section{Dimensional raising map and the class shift}\label{class shift}
In this appendix we discuss the shift of the real AZ or real AZ$+I$ class due to the dimensional raising map, following the formalism of Ref.~\onlinecite{PhysRevB.82.115120}. 

\subsection{Symmetry classes with odd $s$}
First, consider the case where the original Hamiltonian belongs to the chiral symmetric class, that is, the class labeled with an odd number $s$. We take a convention such that the (parity-)time-reversal operator $T$ ($PT$) and the (parity-)particle-hole operator $C$ ($PC$) are commutative. We define the chiral operator $\Gamma$, which is an operator that is proportional to $TC$ or $(PT)(PC)$ and squares to $+\mathbb{I}$, by:
\begin{equation}
\Gamma=
\begin{cases}
TC\ \mathrm{or} \ PTPC \quad(s=1,5), \\
iTC\ \mathrm{or} \ iPTPC\quad(s=3,7).
\end{cases}    
\end{equation}
Then, the commutation relations between $\Gamma$ and $T$ ($PT$), and that between $\Gamma$ and $C$ ($PC$) are
\begin{align}
T\Gamma T^{-1}=C\Gamma C^{-1}=
\begin{cases}
\Gamma\quad(s=1,5),\\
-\Gamma\quad(s=3,7),
\end{cases}
\label{TC_Com}
\end{align}
for real AZ classes and,
\begin{align}
PT\Gamma (PT)^{-1}=PC\Gamma (PC)^{-1}=
\begin{cases}
\Gamma\quad(s=1,5),\\
-\Gamma\quad(s=3,7),
\end{cases}
\label{PTPC_Com}
\end{align}
for real AZ$+I$ classes. 
Since the dimensional raising map breaks the chiral symmetry of the original Hamiltonian, the mapped Hamiltonian cannot have both $T$ and $C$ ($PT$ and $PC$).
(Note that if there are both $T$ and $C$ ($PT$ and $PC$) symmetries, the system possesses the chiral symmetry, which is a product of $T$ and $C$ ($PT$ and $PC$).)

\subsubsection{Real AZ+I}
In the original Hamiltonian, $PT$ and $PC$ do not change $\mathbf{k}\in S^d$. 
Assuming that the same $PT$ and $PC$ do not change $\theta$, the question is whether they are commutative or anticommutative with the mapped Hamiltonian $H^{(d+1)}_{\mathrm{nc}}(\mathbf{k},\theta)$.
The commutation relation between $PT$ and $H^{(d+1)}_{\mathrm{nc}}$, and that between $PC$ and $H^{(d+1)}_{\mathrm{nc}}$, can be calculated from Eq.~(\ref{Hnc_d+1}) as:
\begin{align}
PTH^{(d+1)}_{\mathrm{nc}}(\mathbf{k},\theta)(PT)^{-1}
=&\sin\theta H^{(d)}_{\mathrm{c}}(\mathbf{k})\notag\\
&\quad+\cos\theta PT\Gamma (PT)^{-1},
\\
PCH^{(d+1)}_{\mathrm{nc}}(\mathbf{k},\theta)(PC)^{-1}
=&-\sin\theta H^{(d)}_{\mathrm{c}}(\mathbf{k})\notag \\
&\quad+\cos\theta PC\Gamma (PC)^{-1}.
\end{align}
Using Eq.~(\ref{PTPC_Com}), it is shown that $PT$ is conserved when $s=1,5$, and $PC$ is conserved when $s=3,7$. When we check the class shift for each symmetry class $s$, we can confirm that the shift is from $s$ to $s-1$ (mod $8$).

\subsubsection{Real AZ}
As a comparison, we briefly summarize the case for real AZ classes. In the original Hamiltonian, $T$ and $C$ move $\mathbf{k}\in S^{d}$ to $-\mathbf{k}\in S^{d}$. In the mapped Hamiltonian, $T$ and $C$, if preserved, move $(\mathbf{k},\theta)\in S^{d+1}$ to the point $(-\mathbf{k},\pi-\theta)\in S^{d+1}$. Then the commutation relation between $T$ and $H^{(d+1)}_{\mathrm{nc}}$, and that between $C$ and $H^{(d+1)}_{\mathrm{nc}}$, can be calculated from Eq.~(\ref{Hnc_d+1}) as:
\begin{align}
TH^{(d+1)}_{\mathrm{nc}}(\mathbf{k},\theta)T^{-1}
=&\sin(\pi-\theta) H^{(d)}_{\mathrm{c}}(-\mathbf{k})
\notag \\&\quad
-\cos(\pi-\theta) T\Gamma T^{-1},
\\
CH^{(d+1)}_{\mathrm{nc}}(\mathbf{k},\theta)C^{-1}
=&-\sin(\pi-\theta) H^{(d)}_{\mathrm{c}}(-\mathbf{k})
\notag \\&\quad
-\cos(\pi-\theta) C\Gamma C^{-1}.
\end{align}
Using Eq.~(\ref{TC_Com}), we can see that $C$ is conserved when $s=1,5$, and $T$ is conserved when $s=3,7$. Then, we can confirm that the class shift is from $s$ to $s+1$ (mod $8$).

\subsection{Symmetry classes with even $s$}
Next, consider symmetry classes without chiral symmetry, that is, classes whose label $s$ is an even number. These symmetry classes only have antiunitary symmetries of either $T$ ($PT$) or $C$ ($PC$), and the dimensional raising map is chosen to preserve these symmetries. 

\subsubsection{Real AZ+I}
For the mapped Hamiltonian (Eq.~(\ref{Hc_d+1})), the original symmetries $PT$ and $PC$ are extended to the forms $PT\otimes\mathbb{I}$ and $PC\otimes\mathbb{I}$, respectively. 
In order for these symmetries to be preserved in the mapped Hamiltonian, it is sufficient that $PT\otimes\mathbb{I}$ and $\mathbb{I}\otimes\tau_a$ are commutative, and $PC\otimes\mathbb{I}$ and $\mathbb{I}\otimes\tau_a$ are anti-commutative. That is, it is sufficient to set $\tau_a=\tau_x$ for $PT$ and $\tau_a=\tau_y$ for $PC$.

As mentioned in the main text, the mapped Hamiltonian has an additional chiral symmetry $\Gamma=\mathbb{I}\otimes i\tau_z\tau_a$, which causes a shift in the symmetry class.

For example, if the original Hamiltonian only has the $PT$ symmetry, the mapped Hamiltonian has an additional symmetry constructed by the product of $PT$ and $\Gamma$, which behaves as a $PC$ symmetry:  $PC \propto\Gamma\cdot(PT\otimes\mathbb{I})$. Since the square of this newly defined $PC$ operator does not depend on the phase factor, $(PC)^2=(PT\otimes(-\tau_y))^2=-(PT)^2\otimes\mathbb{I}$. Therefore, classes with $s=0$ and $4$ are mapped to classes with $s=7$ and $3$, respectively.

On the other hand, if the original Hamiltonian only has the $PC$ symmetry, the mapped Hamiltonian has an additional symmetry constructed by the product of $PC$ and $\Gamma$, which behaves as a $PT$ symmetry:  $PT \propto\Gamma\cdot(PC\otimes\mathbb{I})$. Its square is $(PT)^2=(PC\otimes\tau_x)^2=(PC)^2\otimes\mathbb{I}$. Therefore, classes with $s=2$ and $6$ are mapped to classes with $s=1$ and $5$, respectively.

In summary, the symmetry class shifts from $s$ to $s-1$ (mod $8$). 

\subsubsection{Real AZ}
As a comparison, we also discuss the case for real AZ classes. The extended symmetries $T\otimes\mathbb{I}$ and  $C\otimes\mathbb{I}$ act on the mapped Hamiltonian respectively as follows:
\begin{align}
&(T\otimes\mathbb{I})H^{(d+1)}_{\mathrm{c}}(\mathbf{k},\theta)(T\otimes\mathbb{I})^{-1}
\notag \\
=&\sin(\pi-\theta) H^{(d)}_{\mathrm{c}}(-\mathbf{k})\otimes\tau_z
\notag \\
&\ -\cos(\pi-\theta) (T\otimes\mathbb{I})(\mathbb{I}\otimes\tau_a) (T\otimes\mathbb{I})^{-1},
\\
&(C\otimes\mathbb{I})H^{(d+1)}_{\mathrm{c}}(\mathbf{k},\theta)(C\otimes\mathbb{I})^{-1}
\notag \\
=&-\sin(\pi-\theta) H^{(d)}_{\mathrm{c}}(-\mathbf{k})\otimes\tau_z
\notag \\
&\ -\cos(\pi-\theta) (C\otimes\mathbb{I})(\mathbb{I}\otimes\tau_a) (C\otimes\mathbb{I})^{-1}.
\end{align}
Therefore, in order for $T\otimes\mathbb{I}$ ($C\otimes\mathbb{I}$) to be the (anti-)symmetry of the mapped Hamiltonian, it is sufficient to set $\tau_a=\tau_y$ ($\tau_a=\tau_x$). 

If the original Hamiltonian only has $T$, the mapped Hamiltonian additionally has $C_{\mathrm{add}}\propto i\tau_z\tau_y\cdot(T\otimes\mathbb{I})=T\otimes\tau_x$. Then, the square of $C_{\mathrm{add}}$ is equal to $T^2\otimes\mathbb{I}$, and the classes with $s=0$ and $4$ are mapped to the classes with $s=1$ and $5$, respectively. On the other hand, if the original Hamiltonian only has $C$, the mapped Hamiltonian additionally has $T_{\mathrm{add}}\propto i\tau_z\tau_x\cdot(C\otimes\mathbb{I})=C\otimes(-\tau_y)$. Then, the square value of $T_{\mathrm{add}}$ is equal to $-C^2\otimes\mathbb{I}$, and the classes with $s=2$ and $6$ are mapped to the classes with $s=3$ and $7$, respectively. In summary, the symmetry class shift is from $s$ to $s+1$ (mod $8$).

\section{Schmidt decomposition of the ground state}\label{Deriv_GS_vac}
Here we derive Eq.~(\ref{eq:gs}), which is the the Schmidt decomposition of the ground state. For non-interacting fermions, the ground state $\ket{\text{GS}}$ of $\hat{H}$ is given by the Slater determinant of the eigenstates of the correlation matrix $C^{\text{cor}}$ with the eigenvalue 1. Here, following the discussion given in Appendix E of the paper [\onlinecite{PhysRevB.85.165120}], we express the eigenstates of $C^{\text{cor}}$ in terms of the eigenstates of the submatrices $C_A$ and $C_B$ to derive the desired expression.

First, since $(C^{\text{cor}})^2=C^{\text{cor}}$, the following holds:
\begin{align}
C_B C_{BA} &= C_{BA}(\mathbb{I}-C_A),\label{CB_CBA}
\\
C_{AB}C_{BA} &= C_A(\mathbb{I}-C_A).
\label{CAB_CBA}
\end{align}
Then, for eigenvalues and normalized eigenstates of $C_A$, $C_A\ket{f_l^{A}}=\tilde{\xi}_l\ket{f_l^{A}}$, the following holds:
\begin{align}
C_B C_{BA}\ket{f_l^{A}}&=(1-\tilde{\xi}_l)C_{BA}\ket{f_l^{A}},\label{B3}
\\
C_{AB}C_{BA}\ket{f_l^{A}}&=\tilde{\xi}_l(1-\tilde{\xi}_l)\ket{f_l^{A}}.\label{B4}
\end{align}
where $0\leq\tilde{\xi_l}\leq1$. 
In particular, if we denote $\tilde{\xi_l}\neq$0, 1 as $\xi_l$, one can show that the states
\begin{align}
\ket{F_l^{(1)}}&\equiv
\begin{pmatrix}
\sqrt{\xi_l}\ket{f_l^{A}}\\
\frac{C_{BA}\ket{f_l^{A}}}{\sqrt{\xi_l}}
\end{pmatrix},&
\ket{F_l^{(0)}}&\equiv
\begin{pmatrix}
\sqrt{1-\xi_l}\ket{f_l^{A}}\\
-\frac{C_{BA}\ket{f_l^{A}}}{\sqrt{1-\xi_l}}
\end{pmatrix}
\end{align}
are the normalized eigenstates of $C^{\text{cor}}$ with the eigenvalues 1 and 0, respectively.
Since the state $\ket{f_l^B}\equiv C_{BA}\ket{f_l^A}/{\sqrt{\xi_l(1-\xi_l)}}$ is a normalized eigenstate of $C_B$ with the eigenvalue $1 - \xi_l$, we note that these eigenstates of $C^{\text{cor}}$ can also be written as
\begin{align}
\ket{F_l^{(1)}}&\equiv
\begin{pmatrix}
\sqrt{\xi_l}\ket{f_l^{A}}\\
\sqrt{1 - \xi_l} |f_l^B\rangle
\end{pmatrix},&
\ket{F_l^{(0)}}&\equiv
\begin{pmatrix}
\sqrt{1-\xi_l}\ket{f_l^{A}}\\
-\sqrt{\xi_l}|f_l^B\rangle
\end{pmatrix}.
\end{align}

Next, suppose $\ket{g_l^{A}}$ and $\ket{g_l^{B}}$ are the eigenstates corresponding to the eigenvalue 1 of $C_A$ and $C_B$, respectively. Then, the states
\begin{align}
\ket{G_l^{(A)}} &\equiv \begin{pmatrix}
\ket{g_l^{A}}\\
\mathbf{0}
\end{pmatrix},
&
\ket{G_l^{(B)}} &\equiv \begin{pmatrix}
\mathbf{0}
\\
\ket{g_l^{B}}
\end{pmatrix}
\end{align}
are the normalized eigenstates of $C^{\text{cor}}$ with the eigenvalues 1.

Similarly, if $\ket{h_l^{A}}$ and $\ket{h_l^{B}}$ are the eigenstates corresponding to the eigenvalue 0 of $C_A$ and $C_B$, respectively, the states 
\begin{align}
\ket{H_l^{(A)}} &\equiv \begin{pmatrix}
\ket{h_l^{A}}\\
\mathbf{0}
\end{pmatrix},
&
\ket{H_l^{(B)}} &\equiv \begin{pmatrix}
\mathbf{0} \\
\ket{h_l^{B}}
\end{pmatrix}
\end{align}
are the normalized eigenstates of $C^{\text{cor}}$ with the eigenvalue 0.

The six types of eigenstates of $C^{\text{cor}}$ mentioned above are orthogonal to each other and form a basis for the space on which the matrix $C^{\text{cor}}$ acts. Therefore the eigenstates of $C^{\text{cor}}$ with an eigenvalue of 1 are exhausted by $\ket{F_l^{(1)}}$, $\ket{G_l^{(A)}}$ and $\ket{G_l^{(B)}}$. Writing the $i$-th component of $|f_l^{A}\rangle$, $|f_l^{B}\rangle$, $\ket{G_l^{(A)}}$ and $\ket{G_l^{(B)}}$ by $(f_l^A)_i$, $(f_l^B)_i$, $(g_l^{A})_i$, and $(g_l^{B})_i$ respectively, the ground state $\ket{\text{GS}}$ is given by
\begin{align}
\ket{\text{GS}}=&\prod_{n_A}\Bigl{(}\sum_{i\in A}(g^A_{n_A})_i c^{\dagger}_i\Bigr{)}\times
\prod_{n_B}\Bigl{(}\sum_{j\in B}(g^B_{n_B})_j c^{\dagger}_j\Bigr{)}
\notag \\
&\times \prod^{N}_{l=1}\Bigl{(}\sqrt{\xi_l}\sum_{i\in A}(f^A_l)_i c^{\dagger}_i
+
\sqrt{1 - \xi_l}\sum_{j\in B}(f^B_l)_j c^{\dagger}_j
\Bigr{)}
\notag \\
&\times\ket{\text{0}},
\label{GS_vac_fA}
\end{align}
where the products over $n_A$ and $n_B$ are taken over all the eigenstates of $C^{\text{cor}}$ with the eigenvalue 1.
The product over $l$ is taken to cover all the eigenvalues $\xi_l \neq 0$ or $1$; we write the number of such eigenvalues as $N$. Expanding the product over $l$, the third factor in Eq.~(\ref{GS_vac_fA}) becomes as follows:
\begin{align}
&\prod^{N}_{l=1}\Bigl{(}\sqrt{\xi_l}\sum_{i\in A}(f^A_l)_i c^{\dagger}_i
+
\sqrt{1 - \xi_l}\sum_{j\in B}(f^B_l)_j c^{\dagger}_j
\Bigr{)}
\notag \\
=&
\sum_{\{n_l\}}
\prod_{l=1}^{N}\Big{\{}\xi_l^{\frac{1}{2}}\sum_{i\in A}(f^A_l)_i c^{\dagger}_i\Big{\}}^{n_l}
\Big{\{}(1-\xi_l)^{\frac{1}{2}}\sum_{j\in B}(f^B_l)_j c^{\dagger}_j\Big{\}}^{1-n_l},
\end{align}
where $n_l \in \{ 0, 1\}$; the set $\{ n_l \}$ takes its value in $\{0, 1\}^N$and the summation over $\{n_l\}$ is for all possible values of $\{ n_l \} \in \{0, 1\}^N$.
As a result, Eq.~(\ref{GS_vac_fA}) can be rewritten as follows:
\begin{align}
\ket{\text{GS}}
=&\sum_{\{n_l\}}(-1)^{S(\{n_l\})}
\prod_l\xi_l^{\frac{n_l}{2}}
(1-\xi_l)^{\frac{1-n_l}{2}}
\ket{\alpha_{\{n_l\}}^{A}}\otimes\ket{\alpha_{\{n_l\}}^{B}}, \label{eq:App_B_GS}
\\
\ket{\alpha_{\{n_l\}}^{A}}&=
\prod_{n_A}\Bigl{(}\sum_{i\in A}(g^A_{n_A})_i c^{\dagger}_i\Bigr{)}\prod_l\Bigl{(}\sum_{i\in A}(f^A_l)_i c^{\dagger}_i\Bigr{)}^{n_l}\ket{\text{0}^{A}},
\\
\ket{\alpha_{\{n_l\}}^{B}}&=\prod_{n_B}\Bigl{(}\sum_{i\in B}(g^B_{n_B})_i c^{\dagger}_i\Bigr{)}\prod_l
\Bigl{(}\sum_{j\in B}(f^B_l)_j c^{\dagger}_j
\Bigr{)}^{1-n_l}\ket{\text{0}^{B}},
\end{align}
where $S(\{n_l\})$ corresponds to the sign factor that can arise from the commutation of creation operators. Here, Eq.~(\ref{eq:App_B_GS}) is exactly the Eq.~(\ref{eq:gs}) we intended to derive in this appendix.

\section{General properties of $\Xi$}
\label{General_Xi}
Here, we derive some important properties of $\Xi$.
We start from the general flattened Hamiltonian of Eq.~(\ref{H_flat_general}) written in the following block form:
\begin{align}
    H_{\text{flat}}=\begin{pmatrix}
        H_A & H_{AB} \\
        H_{BA} & H_B 
    \end{pmatrix}.
    \label{H_flat_block}
\end{align}
Since all the eigenvalues of $H_{\text{flat}}$ are either $\pm1/2$, it follows that $(H_{\text{flat}})^2=\mathbb{I}/4$. Consequently, the following equations hold:
\begin{align}
    H_A^2+H_{AB}H_{BA}=\frac{\mathbb{I}_A}{4}, \label{H_flat_square_A} \\
    H_{BA}H_{A}+H_{B}H_{BA}=\mathbf{0}. \label{H_flat_square_BA} 
\end{align}
From Eq.~(\ref{H_flat_square_BA}), if $\ket{\psi_n}$ is an eigenstate of $H_A$ with the eigenvalue $\epsilon_n$, the following holds:
\begin{align}
    H_{B}H_{BA}\ket{\psi_n}=-\epsilon_n H_{BA}\ket{\psi_n}. \label{HBA_eigen}
\end{align}
The norm of $H_{BA}\ket{\psi_n}$ is determined from Eq.~(\ref{H_flat_square_A}) as follows:
\begin{align}
    \bra{\psi_n}[H_{BA}]^{\dagger}H_{BA}\ket{\psi_n}&=\bra{\psi_n}\mathbb{I}_A/4-H_{A}^{2}\ket{\psi_n} \notag \\
    &=\Big{(}\frac{1}{4}-\epsilon_n^2\Big{)}.\label{H_psi_norm}
\end{align}
Here, we used $H_{AB}=H_{BA}^{\dagger}$, which comes from the Hermiticity of $H_{\text{flat}}$. 
From Eqs.~(\ref{HBA_eigen}) and (\ref{H_psi_norm}), the state $\ket{\phi_n}$ defined as
\begin{align}
    \ket{\phi_n}\equiv-\frac{H_{BA}}{\sqrt{1/4-\epsilon_n^2}}\ket{\psi_n},
    \label{phi_n_def}
\end{align}
is a normalized eigenstate of $H_B$ with the eigenvalue $-\epsilon_n$, as long as $\epsilon_n\neq\pm1/2$. Here, the negative sign is introduced to be consistent with the definition used in a previous work\cite{PhysRevB.108.075129}. 
It is also possible to construct the eigenstates of $H_A$ by applying $H_{AB}$ to the eigenstates of $H_B$. In particular, from Eqs.~(\ref{H_flat_square_A}) and (\ref{phi_n_def}), the following holds:
\begin{align}
    \ket{\psi_n}=-\frac{H_{AB}}{\sqrt{1/4-\epsilon_n^2}}\ket{\phi_n}.
\end{align}
Then, we can formally define the matrix $\Xi$ as follows:
\begin{align}
    \Xi&\equiv
    \begin{pmatrix}
    \mathbf{0} & \Xi_{AB} \\
    \Xi_{BA} & \mathbf{0}
    \end{pmatrix} \\ 
    &\equiv\sum_{n (\epsilon_n\neq\pm\frac{1}{2})}\frac{1}{\sqrt{1/4-\epsilon_n^2}}
    \begin{pmatrix}
    \mathbf{0} & H_{AB}\ket{\phi_n}\bra{\phi_n} \\
    H_{BA}\ket{\psi_n}\bra{\psi_n} & \mathbf{0}
    \end{pmatrix} \label{Xi_def_Hpsi} \\
    &=\sum_{n (\epsilon_n\neq\pm\frac{1}{2})}
    \begin{pmatrix}
    \mathbf{0} & \ket{\psi_n}\bra{\phi_n} \\
    \ket{\phi_n}\bra{\psi_n} & \mathbf{0}
    \end{pmatrix}. \label{Xi_psi_phi}
\end{align}
The operation $\Xi$ defined here maps the eigenstate $\ket{\psi_n}$ of $H_A$, which is in $A$ ($\simeq[\ket{\psi_n}^{T},\mathbf{0}_B^{T}]^{T}$), to the eigenstate $\ket{\phi_n}=\Xi_{BA}\ket{\psi_n}$ of $H_B$, which is in $B$ ($\simeq[\mathbf{0}_A^{T},\ket{\phi_n}^{T}]^{T}$). 

The square of $\Xi$ is
\begin{align}
    \Xi^2=\sum_{n (\epsilon_n\neq\pm\frac{1}{2})}
    \begin{pmatrix}
    \ket{\psi_n}\bra{\psi_n} & \mathbf{0} \\
    \mathbf{0} & \ket{\phi_n}\bra{\phi_n} 
    \end{pmatrix}.
\end{align}
In each subspace $A$ and $B$, $\Xi^2$ is a projection to the eigenspace where the eigenvalues of $H_A$ and $H_B$ are $\epsilon\neq \pm 1/2$, respectively. 

\section{Commutation relations between the antisymmetry $\Xi$ and other three symmetries}
\label{commutation relation}
In this subsection, we discuss the commutation relations between $\Xi$ and other three symmetries, $PT$, $PC$ and $\Gamma$. 
The Hamiltonian $H_{\mathrm{flat}}(\mathbf{k}_{\perp})$ possesses symmetries such as $PT$, $PC$, and $\Gamma$, which are also present in the original system. The symmetries $PT$ and $PC$ swap regions $A$ and $B$, while the chiral symmetry $\Gamma$ does not swap the regions.
Then, these symmetries can be written in the following block diagonal or anti-diagonal form:
\begin{align}
    (PT)&=\begin{pmatrix}
        \mathbf{0} & U_T \\
        V_T & \mathbf{0} 
    \end{pmatrix}K, \label{PT_block}\\
    (PC)&=\begin{pmatrix}
        \mathbf{0} & U_C \\
        V_C & \mathbf{0} 
    \end{pmatrix}K, \label{PC_block}\\
    \Gamma&=\begin{pmatrix}
        U_\Gamma & \mathbf{0} \\
        \mathbf{0} & V_\Gamma
    \end{pmatrix}, \label{S_diag}
\end{align}
where $U_T$, $V_T$, $U_C$, $V_C$, $U_\Gamma$ and $V_\Gamma$ are unitary matrices, and $K$ is the complex conjugation operation. From the commutation relations $[H_{\mathrm{flat}},PT]=\{H_{\mathrm{flat}},PC\}=\{H_{\mathrm{flat}},\Gamma\}=0$, the following relations hold:
\begin{align}
    \begin{pmatrix}
        U_T[H_{BA}]^{*} & U_T[H_{B}]^{*} \\
        V_T[H_{A}]^{*} & V_T[H_{AB}]^{*}
    \end{pmatrix}
    &=
    \begin{pmatrix}
        H_{AB}V_T & H_{A}U_T \\
        H_{B}V_T & H_{BA}U_T
    \end{pmatrix}, \label{UT_H} \\
    \begin{pmatrix}
        U_C[H_{BA}]^{*} & U_C[H_{B}]^{*} \\
        V_C[H_{A}]^{*} & V_C[H_{AB}]^{*}
    \end{pmatrix}
    &=-
    \begin{pmatrix}
        H_{AB}V_C & H_{A}U_C \\
        H_{B}V_C & H_{BA}U_C
    \end{pmatrix}, \label{UC_H} \\
    \begin{pmatrix}
        U_\Gamma H_{A} & U_\Gamma H_{AB} \\
        V_\Gamma H_{BA} & V_\Gamma H_{B}
    \end{pmatrix}
    &=-
    \begin{pmatrix}
        H_{A}U_\Gamma & H_{AB}V_\Gamma \\
        H_{BA}U_\Gamma & H_{B}V_\Gamma
    \end{pmatrix}. \label{US_H} 
\end{align}
Here, we abbreviate $H_{\mathrm{flat},\alpha}(\mathbf{k}_{\perp})$ by $H_{\alpha}$ ($\alpha=A, B, AB, BA$). 
From these equations, it can be seen that the $PT$ and $PC$ symmetries map the eigenstates of $H_A$ to those of $H_B$, and vice versa. 
For example, if $\ket{\phi_n}$ is the eigenstate of $H_B$ with the eigenvalue $\epsilon_n$, $U_T\ket{\phi_n^{*}}$ and $U_C\ket{\phi_n^{*}}$ are the eigenstates of $H_A$:
\begin{align}
H_A U_T\ket{\phi_n^{*}}
&=U_T[H_B]^{*}\ket{\phi_n^{*}} \notag \\
&=\epsilon_n U_T\ket{\phi_n^{*}}, 
\\
H_A U_C\ket{\phi_n^{*}}
&=-U_C[H_B]^{*}\ket{\phi_n^{*}} \notag \\
&=-\epsilon_n U_C\ket{\phi_n^{*}}. 
\end{align}
Similarly, if $\ket{\psi_n}$ is the eigenstate of $H_A$ with the eigenvalue $\epsilon_n$, $V_T\ket{\psi_n^{*}}$ and $V_C\ket{\psi_n^{*}}$ are the eigenstates of $H_B$. 

Now we discuss the commutation relations between $\Xi$ and the three symmetries $PT$, $PC$, and $\Gamma$.

\subsection{Commutation relation between $\Xi$ and $PT$}
From Eqs.~(\ref{Xi_def_Hpsi}), (\ref{PT_block}) and (\ref{UT_H}),
\begin{align}
    &PT\Xi \notag \\
    =&\sum_{n (\epsilon_n\neq\pm\frac{1}{2})}\frac{1}{\sqrt{\frac{1}{4}-\epsilon_n^2}}
    \begin{pmatrix}
    U_TH_{BA}^{*}\ket{\psi_n^{*}}\bra{\psi_n^{*}} & \mathbf{0} \\
    \mathbf{0} & V_TH_{AB}^{*}\ket{\phi_n^{*}}\bra{\phi_n^{*}} 
    \end{pmatrix}K \notag \\
    =&\sum_{n (\epsilon_n\neq\pm\frac{1}{2})}\frac{1}{\sqrt{\frac{1}{4}-\epsilon_n^2}}
    \begin{pmatrix}
    H_{AB}V_T\ket{\psi_n^{*}}\bra{\psi_n^{*}} & \mathbf{0} \\
    \mathbf{0} & H_{BA}U_T\ket{\phi_n^{*}}\bra{\phi_n^{*}} 
    \end{pmatrix}K. \label{PTXi_1}
\end{align}
Here, for the eigenstate $\ket{\psi_n}$ of $H_A$ with eigenvalue $\epsilon_n$ from Eq.~(\ref{UT_H}), the following hold:
\begin{align}
H_B(V_T\ket{\psi_n^{*}})=V_TH_A^{*}\ket{\psi_n^{*}}=\epsilon_nV_T\ket{\psi_n^{*}}.
\end{align}
This means that $V_T\ket{\psi_n^{*}}$ is an eigenstate of $H_B$. Since, $V_TK$ do not change the absolute value of the inner product, $\{V_T\ket{\psi_n^{*}}\}$ forms another basis of the eigenspace of $H_B$ associated with the eigenvalues $\epsilon_n\neq \pm 1/2$. Then, the following equation between the two projectors holds:
\begin{align}
    \sum_{n (\epsilon_n\neq\pm\frac{1}{2})}V_T\ket{\psi_n^{*}}\bra{\psi_n^{*}}V_T^{\dagger}=\sum_{n (\epsilon_n\neq\pm\frac{1}{2})}\ket{\phi_n}\bra{\phi_n}.
    \label{VTK_proj}
\end{align}
Similarly, from Eq.~(\ref{UT_H}), $U_T\ket{\phi_n^{*}}$ is an eigenstate of $H_A$, and the following holds:
\begin{align}
    \sum_{n (\epsilon_n\neq\pm\frac{1}{2})}U_T\ket{\phi_n^{*}}\bra{\phi_n^{*}}U_T^{\dagger}=\sum_{n (\epsilon_n\neq\pm\frac{1}{2})}\ket{\psi_n}\bra{\psi_n}.
    \label{UTK_proj}
\end{align}
By using Eqs.~(\ref{VTK_proj}) and (\ref{UTK_proj}), Eq.~(\ref{PTXi_1}) can be rewritten as:
\begin{align}
    &PT\Xi = \notag \\
    &\sum_{n (\epsilon_n\neq\pm\frac{1}{2})}\frac{1}{\sqrt{\frac{1}{4}-\epsilon_n^2}}
    \begin{pmatrix}
    H_{AB}\ket{\phi_n}\bra{\phi_n}V_T & \mathbf{0} \\
    \mathbf{0} & H_{BA}\ket{\psi_n}\bra{\psi_n}U_T 
    \end{pmatrix}K \notag \\
    &=\Xi PT, 
\end{align}
which shows that $PT$ and $\Xi$ commute.

\subsection{Commutation relation between $\Xi$ and $PC$}
From Eqs.~(\ref{Xi_def_Hpsi}), (\ref{PC_block}) and (\ref{UC_H}),
\begin{align}
    &PC\Xi = \notag \\
    &\sum_{n (\epsilon_n\neq\pm\frac{1}{2})}\frac{1}{\sqrt{\frac{1}{4}-\epsilon_n^2}}
    \begin{pmatrix}
    U_CH_{BA}^{*}\ket{\psi_n^{*}}\bra{\psi_n^{*}} & \mathbf{0} \\
    \mathbf{0} & V_CH_{AB}^{*}\ket{\phi_n^{*}}\bra{\phi_n^{*}} 
    \end{pmatrix}K \notag \\
    &= \notag \\
    &\sum_{n (\epsilon_n\neq\pm\frac{1}{2})}\frac{(-1)}{\sqrt{\frac{1}{4}-\epsilon_n^2}}
    \begin{pmatrix}
    H_{AB}V_C\ket{\psi_n^{*}}\bra{\psi_n^{*}} & \mathbf{0} \\
    \mathbf{0} & H_{BA}U_C\ket{\phi_n^{*}}\bra{\phi_n^{*}} 
    \end{pmatrix}K. \label{PCXi_1}
\end{align}
Here, for the eigenstate $\ket{\psi_n}$ of $H_A$ with eigenvalue $\epsilon_n$, from Eq.~(\ref{UC_H}), $V_C\ket{\psi_n^{*}}$ is an eigenstate of $H_B$ with the eigenvalue $-\epsilon_n$. Similarly, $U_C\ket{\phi_n^{*}}$ is an eigenstate of $H_A$. Therefore, similar to Eqs.~(\ref{VTK_proj}) and (\ref{UTK_proj}), the following equations hold:
\begin{align}
    \sum_{n (\epsilon_n\neq\pm\frac{1}{2})}V_C\ket{\psi_n^{*}}\bra{\psi_n^{*}}V_C^{\dagger}&=\sum_{n (\epsilon_n\neq\pm\frac{1}{2})}\ket{\phi_n}\bra{\phi_n},
    \label{VCK_proj} \\
    \sum_{n (\epsilon_n\neq\pm\frac{1}{2})}U_C\ket{\phi_n^{*}}\bra{\phi_n^{*}}U_C^{\dagger}&=\sum_{n (\epsilon_n\neq\pm\frac{1}{2})}\ket{\psi_n}\bra{\psi_n}.
    \label{UCK_proj}
\end{align}
Then, using Eqs.~(\ref{VCK_proj}) and (\ref{UCK_proj}), Eq.~(\ref{PCXi_1}) can be rewritten as:
\begin{align}
    &PC\Xi =\notag \\
    &\sum_{n (\epsilon_n\neq\pm\frac{1}{2})}\frac{(-1)}{\sqrt{\frac{1}{4}-\epsilon_n^2}}
    \begin{pmatrix}
    H_{AB}\ket{\phi_n}\bra{\phi_n}V_C & \mathbf{0} \\
    \mathbf{0} & H_{BA}\ket{\psi_n}\bra{\psi_n}U_C 
    \end{pmatrix}K \notag \\
    &=-\Xi PC,
\end{align}
which shows that $PC$ and $\Xi$ anti-commutate.

\subsection{Commutation relation between $\Xi$ and $\Gamma$}
From Eqs.~(\ref{Xi_def_Hpsi}), (\ref{S_diag}) and (\ref{US_H}),
\begin{align}
    &\Gamma\Xi = \notag \\
    &\sum_{n (\epsilon_n\neq\pm\frac{1}{2})}\frac{1}{\sqrt{\frac{1}{4}-\epsilon_n^2}}
    \begin{pmatrix}
    \mathbf{0} & U_\Gamma H_{AB}\ket{\phi_n}\bra{\phi_n} \\ 
    V_\Gamma H_{BA}\ket{\psi_n}\bra{\psi_n} & \mathbf{0} \end{pmatrix}\notag \\ &= \notag \\
    &\sum_{n (\epsilon_n\neq\pm\frac{1}{2})}\frac{(-1)}{\sqrt{\frac{1}{4}-\epsilon_n^2}}
    \begin{pmatrix}
    \mathbf{0} & H_{AB}V_\Gamma \ket{\phi_n}\bra{\phi_n} \\ 
    H_{BA}U_\Gamma \ket{\psi_n}\bra{\psi_n} & \mathbf{0} \end{pmatrix}. \label{SXi_1}
\end{align}
Here, from Eq.~(\ref{US_H}), the states $U_\Gamma \ket{\psi_n}$ and $V_\Gamma \ket{\phi_n}$ are eigenstates of $H_A$ and $H_B$, respectively. 
Therefore, the following equations hold:
\begin{align}
    \sum_{n (\epsilon_n\neq\pm\frac{1}{2})}U_\Gamma \ket{\psi_n}\bra{\psi_n}U_\Gamma^{\dagger}&=\sum_{n (\epsilon_n\neq\pm\frac{1}{2})}\ket{\psi_n}\bra{\psi_n},
    \label{US_proj} \\
    \sum_{n (\epsilon_n\neq\pm\frac{1}{2})}V_\Gamma \ket{\phi_n}\bra{\phi_n}V_\Gamma^{\dagger}&=\sum_{n (\epsilon_n\neq\pm\frac{1}{2})}\ket{\phi_n}\bra{\phi_n}.
    \label{VS_proj}
\end{align}
Then, using Eqs.~(\ref{US_proj}) and (\ref{VS_proj}), Eq.~(\ref{SXi_1}) can be rewritten as:
\begin{align}
    &\Gamma \Xi \notag \\
    =&\sum_{n (\epsilon_n\neq\pm\frac{1}{2})}\frac{(-1)}{\sqrt{\frac{1}{4}-\epsilon_n^2}}
    \begin{pmatrix}
    \mathbf{0} & H_{AB}\ket{\phi_n}\bra{\phi_n}V_\Gamma \\ 
    H_{BA}\ket{\psi_n}\bra{\psi_n}U_\Gamma & \mathbf{0} \end{pmatrix} \notag \\
    =&-\Xi \Gamma,
\end{align}
which shows that $\Gamma$ and $\Xi$ anti-commute.

\section{Berry phase in class AI$^{\prime}$}\label{AI_Berry}
The topological phases of one-dimensional class AI$^\prime$ systems are classified by the $\mathbb{Z}_2$-valued topological number. If the $PT$ symmetry is represented by $PT=K$ with the complex conjugate operation $K$, then the Berry phase is quantized to $0$ or $\pi$, and the $\mathbb{Z}_2$ topological number is equivalent to the Berry phase. However, in general, $PT$ symmetry is expressed as $PT=U(\mathbf{k})K$ with a unitary matrix $U(\mathbf{k})$, and if $U(\mathbf{k})$ depends on the wave number, the Berry phase itself is not quantized in general. Instead, a modified Berry phase with a correction term is quantized, as we show below. 
It is worth mentioning that there are several previous works on the $k$-dependence of inversion operations and the non-quantized Berry phase, specifically within the context of non-centered inversion operations \cite{PhysRevB.100.041104,PhysRevB.109.195116}.

We consider a class AI$^{\prime}$ Hamiltonian $H(\mathbf{k})$, which has $PT$ symmetry $[PT,H(\mathbf{k})]$ (i.e., $UH^{*}(\mathbf{k})=H(\mathbf{k})U$). We assume that $PT=U(\mathbf{k})K$ and $(PT)^2=\mathbb{I}$.
For simplicity, we focus on one band without band degeneracy and consider the Berry phase for that band.
Then, for the normalized eigenstate $\ket{u_n(\mathbf{k})}$ of $H(\mathbf{k})$, the following holds:
\begin{align}
    PT\ket{u_n(\mathbf{k})}
    =U\ket{u_n^{*}(\mathbf{k})}
    =e^{i\phi_n(\mathbf{k})}\ket{u_n(\mathbf{k})},
\end{align}
\begin{align}
    \ket{u_n(\mathbf{k})}=e^{-i\phi_n(\mathbf{k})}U\ket{u_n^{*}(\mathbf{k})},
\end{align}
\begin{align}
    \langle u_n\ket{\nabla_{\mathbf{k}}u_n}
    =&(-i\nabla_{\mathbf{k}}\phi_n)+\bra{u_n^{*}}U^{\dagger}(\nabla_{\mathbf{k}}U)\ket{u^{*}_n} \notag \\
    &\quad +\langle u^{*}_n\ket{\nabla_{\mathbf{k}}u^{*}_n}.
\end{align}
Here, the third term is rewriten as $\langle u^{*}_n\ket{\nabla_{\mathbf{k}}u^{*}_n}=-\langle u_n\ket{\nabla_{\mathbf{k}}u_n}$. 
Then the Berry phase (times 2) along a path $C$ in $\mathbf{k}$-space can be written as:
\begin{align}
    &2i\oint_C \langle u_n\ket{\nabla_{\mathbf{k}}u_n}\cdot d\mathbf{k}
    \notag \\
    =&\oint_C (\nabla_{\mathbf{k}}\phi_n)\cdot d\mathbf{k}+i\oint_C \bra{u_n^{*}}U^{\dagger}(\nabla_{\mathbf{k}}U)\ket{u^{*}_n}\cdot d\mathbf{k}
    \notag \\
    =& 2\pi m+i\oint_C \bra{u_n^{*}}U^{\dagger}(\nabla_{\mathbf{k}}U)\ket{u^{*}_n}\cdot d\mathbf{k} \quad(m\in\mathbb{Z}).
\end{align}
Therefore, instead of the Berry phase itself, the Berry phase with the additional term is quantized:
\begin{align}
    \theta_{\text{Berry}}-\frac{i}{2}\oint_C \bra{u_n^{*}}U^{\dagger}(\nabla_{\mathbf{k}}U)\ket{u^{*}_n}\cdot d\mathbf{k}=m\pi \quad(m\in\mathbb{Z}).
    \label{Berry_quantize_1}
\end{align}
Only if the second term of l.h.s. is zero, the Berry phase $\theta_{\text{Berry}}$ is quantized to 0 or $\pi$. 
For example, if $U$ does not depend on $\mathbf{k}$, the Berry phase $\theta_{\text{Berry}}$ is quantized.

One can alternatively obtain the same topological number as a quantized Berry phase by moving to a basis where the $PT$ symmetry is expressed as a simple complex conjugate operation $K$. 
Below, we give one such unitary transformation of the basis and calculate the Berry phases before and after the transformation.
From $(PT)^2=UU^{*}=\mathbb{I}$, the unitary matrix $U$ is symmetric matrix: $U^{T}=U$. Then, $U$ can be decomposed as $U=X+iY$ using real symmetric matrices $X$ and $Y$. Here, from $UU^{*}=X^2+Y^2-i[X,Y]=\mathbb{I}$, $X$ and $Y$ commute. 
Therefore, $X$ and $Y$ are simultaneously diagonalized by the real orthogonal matrix $W$. This means that $U$ is diagonalized by $W$:
\begin{align}
    W^{T}UW=\begin{pmatrix}
        e^{i\theta_1} & & \\
        & \ddots & \\
        & & e^{i\theta_N}
    \end{pmatrix}
    =e^{i\Theta}.
\end{align}
Therefore,
\begin{align}
    U&=We^{i\Theta}W^{T} \notag \\
    &=V^{\dagger}V^{*},
    \label{U_VV}
\end{align}
where $V=e^{-i\Theta/2}W^{\dagger}$. 
Here, the sign when taking the square root is determined appropriately.
After the basis transformation $V$, the $PT$ symmetry can be represented as:
\begin{align}
    \widetilde{PT}&=V(PT)V^{\dagger}=V(UK)V^{\dagger}
    \notag \\
    &=V(V^{\dagger}V^{*})V^{T}K=K. 
\end{align}
In fact, the transformed Hamiltonian satisfies the following relation:
\begin{align}
    [\widetilde{H(\mathbf{k})}]^{*}
    &=[VH(\mathbf{k})V^{\dagger}]^{*} \notag \\
    &=V^{*}(U^{\dagger}H(\mathbf{k})U)V^{T} \notag \\
    &=V^{*}(V^{T}VH(\mathbf{k})V^{\dagger}V^{*})V^{T} \notag \\
    &=VH(\mathbf{k})V^{\dagger}=\widetilde{H(\mathbf{k})}.
\end{align}
The Berry phase in this basis, $\tilde{\theta}_{\text{Berry}}$, is calculated as follows by using $\widetilde{\ket{u_n}}=V\ket{u_n}$:
\begin{align}
    \tilde{\theta}_{\text{Berry}}
    &=i\oint_C  \bra{u_n}V^{\dagger} \nabla_{\mathbf{k}}(V\ket{u_n})\cdot d\mathbf{k}
    \notag \\
    &=\theta_{\text{Berry}}+
    i\oint_C  \bra{u_n} V^{\dagger}(\nabla_{\mathbf{k}}V) \ket{u_n}\cdot d\mathbf{k}.
    \label{Tilde_Berry}
\end{align}
On the other hand, by using Eq.~(\ref{U_VV}), the integrand of the correction term in Eq.~(\ref{Berry_quantize_1}) is rewritten as the following simpler expression:
\begin{align}
    &\bra{u_n^{*}}U^{\dagger}(\nabla_{\mathbf{k}}U)\ket{u_n^{*}}
    \notag \\
    =&\bra{u_n^{*}}U^{\dagger}(\nabla_{\mathbf{k}}V^{\dagger})V^{*}\ket{u_n^{*}}
      +\bra{u_n^{*}}U^{\dagger}V^{\dagger}(\nabla_{\mathbf{k}}V^{*})\ket{u_n^{*}}
      \notag \\
    =&(U\ket{u_n^{*}})^{\dagger}(\nabla_{\mathbf{k}}V^{\dagger})V(U\ket{u_n^{*}})
      +\bra{u_n^{*}}(V^{T}V)V^{\dagger}(\nabla_{\mathbf{k}}V^{*})\ket{u_n^{*}}
      \notag \\
    =&\bra{u_n}(\nabla_{\mathbf{k}}V^{\dagger})V\ket{u_n}
      +(\bra{u_n}V^{\dagger}(\nabla_{\mathbf{k}}V)\ket{u_n})^{*}
      \notag \\
    =&-2 \bra{u_n}V^{\dagger}(\nabla_{\mathbf{k}}V)\ket{u_n}.
\end{align}
Then, the l.h.s. of Eq.~(\ref{Berry_quantize_1}) becomes as follows:
\begin{align}
    &\theta_{\text{Berry}}-\frac{i}{2}\oint_C \bra{u_n^{*}}U^{\dagger}(\nabla_{\mathbf{k}}U)\ket{u_n}\cdot d\mathbf{k}
    \notag \\
    =&\theta_{\text{Berry}}+i\oint_C  \bra{u_n} V^{\dagger}(\nabla_{\mathbf{k}}V) \ket{u_n}\cdot d\mathbf{k}.
\end{align}
Therefore, from Eq.~(\ref{Tilde_Berry}), the l.h.s. of Eq.~(\ref{Berry_quantize_1}) is equal to $\tilde{\theta}_{\text{Berry}}$. 
Therefore, the quantization of the l.h.s. of Eq.~(\ref{Berry_quantize_1}) is nothing but the well-known Berry phase quantization in the basis with $\widetilde{PT}=K$.

\section{Berry phase in the 3D class C$^{\prime}$ entanglement spectrum}
\label{3D_C_Berry}
In the 3D class C$^{\prime}$ entanglement spectrum, $PC\Xi$ behaves as effective $PT$ symmetry. For the eigenstate $\ket{u_n}$, $PC\Xi$ is represented as follows:
\begin{align}
    PC\Xi\ket{u_n}=\frac{1}{\sqrt{1/4-\epsilon_n^2}}U_CH_{BA}^{*}\ket{u_n^{*}}=R\ket{u_n^{*}}.\label{PCXi_R}
\end{align}
Here, $R$ is generally not a unitary matrix. (This is because, for the eigenstate $\ket{u_n}$ of $H_A$ with the eigenvalue $\epsilon_n=\pm1/2$, $PC\Xi\ket{u_n}=R\ket{u_n^{*}}=\mathbf{0}$, and thus $R$ does not preserve the norm.) Therefore, the discussion in Appendix \ref{AI_Berry} cannot be applied directly.
In this appendix, we will discuss the quantization of the modified Berry phase with a correction term when $R$ is not necessarily a unitary matrix. 

For simplicity, we assume that $\ket{u_n}$ is not degenerate. Then, $PC\Xi\ket{u_n}$ is proportional to $\ket{u_n}$
\begin{align}
    R\ket{u_n^{*}}=e^{i\phi_n}\ket{u_n}.
    \label{Ru_phi_u}
\end{align}
Then,
\begin{align}
    &\langle u_n \ket{\nabla_{\mathbf{k}}u_n}
    \notag \\
    =&\langle u_n^{*}|R^{\dagger}e^{i\phi_n}\nabla_{\mathbf{k}}(e^{-i\phi_n}R\ket{u_n^{*}})
    \notag \\
    =&(-i\nabla_{\mathbf{k}}\phi_n) + \bra{u_n^{*}}R^{\dagger}(\nabla_{\mathbf{k}}R)\ket{u_n^{*}} + \bra{u_n^{*}}R^{\dagger}R\ket{\nabla_{\mathbf{k}}u_n^{*}}.
\end{align}
Here, the third term can be rewritten as:
\begin{align}
    \bra{u_n^{*}}R^{\dagger}R\ket{\nabla_{\mathbf{k}}u_n^{*}}
    =&\frac{\bra{u_n^{*}}H_{BA}^{T}H_{BA}^{*}}{1/4-\epsilon_n^2}\ket{\nabla_{\mathbf{k}}u_n^{*}}
    \notag \\
    =&\langle u_n^{*}\ket{\nabla_{\mathbf{k}}u_n^{*}}
    \notag \\
    =&-\langle u_n \ket{\nabla_{\mathbf{k}}u_n}.
\end{align}
Then, the following holds: 
\begin{align}
    &i\oint_C \langle u_n \ket{\nabla_{\mathbf{k}}u_n}\cdot d\mathbf{k}
    -\frac{i}{2}\oint_C \bra{u_n^{*}}R^{\dagger} (\nabla_{\mathbf{k}}R) \ket{u_n^{*}}\cdot d\mathbf{k}
    \notag \\
    =& \frac{1}{2}\oint_C (\nabla_{\mathbf{k}}\phi_n)\cdot d\mathbf{k}
    \notag \\
    =&m\pi \quad (m\in\mathbb{Z}).
    \label{PCXi_Berry}
\end{align}
Therefore, we use the l.h.s. of Eq.~(\ref{PCXi_Berry}) as the edge topological invariant:
\begin{align}
    &\chi^{(\text{edge})}_{\text{3D-C}^{\prime}}
    \notag \\
    =&\exp\Big{[}
    -\oint_C \langle u_n \ket{\nabla_{\mathbf{k}}u_n}\cdot d\mathbf{k}
    +\frac{1}{2}\oint_C \bra{u_n^{*}}R^{\dagger} (\nabla_{\mathbf{k}}R) \ket{u_n^{*}}\cdot d\mathbf{k}
    \Big{]}
    \notag \\
    =&(-1)^m\quad (m\in\mathbb{Z}).
    \label{edge_3DC}
\end{align}
If $\chi^{(\text{edge})}_{\text{3D-C}^{\prime}}=(-1)$ along some path $C$, it means that a gapless entanglement edge mode exists inside the path $C$.

Finally, we derive a simpler expression for the case where $U_C$ does not depend on the wave number $\mathbf{k}$.
From Eq.~(\ref{Ru_phi_u}), the integrand of the correction term in Eq.~(\ref{edge_3DC}) is rewritten as follows:
\begin{align}
    \bra{u_n^{*}}R^{\dagger}(\nabla_{\mathbf{k}}R)\ket{u_n^{*}}
=\bra{u_n}(\nabla_{\mathbf{k}}R)R^{*}\ket{u_n}.
\label{E7}
\end{align}
Next, using the assumption that $U_C$ is a constant matrix, $(\nabla_{\mathbf{k}}R)R^{*}$ can be rewritten as:
\begin{align}
&(\nabla_{\mathbf{k}}R)R^{*}
\notag \\
=&\nabla_{\mathbf{k}}\Bigg{(}
\frac{U_C H_{\text{BA}}^{*}}{\sqrt{1/4-\epsilon_n^2}}
\Bigg{)}\frac{U_C^{*} H_{\text{BA}}}{\sqrt{1/4-\epsilon_n^2}}
\notag \\
=&\nabla_{\mathbf{k}}\Bigg{(}\frac{H_{\text{AB}}}{\sqrt{1/4-\epsilon_n^2}}
\Bigg{)}\frac{H_{\text{BA}}}{\sqrt{1/4-\epsilon_n^2}}
\notag \\
=&\frac{(\nabla_{\mathbf{k}}H_{\text{AB}})H_{\text{BA}}}{1/4-\epsilon_n^2}
+\frac{\epsilon_n(\nabla_{\mathbf{k}}\epsilon_n)H_{\text{AB}}H_{\text{BA}}}{(1/4-\epsilon_n^2)^2}.
\end{align}
Therefore, r.h.s. of Eq.~(\ref{E7}) is rewritten as follows:
\begin{align}
\bra{u_n}(\nabla_{\mathbf{k}}R)R^{*}\ket{u_n}
=&\frac{\bra{u_n}(\nabla_{\mathbf{k}}H_{\text{AB}})H_{\text{BA}}\ket{u_n}}{1/4-\epsilon_n^2}
+\frac{\epsilon_n(\nabla_{\mathbf{k}}\epsilon_n)}{1/4-\epsilon_n^2}, 
\end{align}
where we used $\bra{u_n}H_{\text{AB}}H_{\text{BA}}\ket{u_n}=\frac{1}{4}-\epsilon_n^2$. 
The second term can be further rewritten as $\nabla_{\mathbf{k}}[-\frac{1}{2}\ln(\frac{1}{4}-\epsilon_n^2)]$, making the circular integral of the second term equal to zero. 
Therefore, the topological invariant in Eq.~(\ref{edge_3DC}) can be rewritten as:
\begin{align}
    &\chi^{(\text{edge})}_{\text{3D-C}^{\prime}}
    \notag \\
    =&\exp\Big{[}
    -\oint_C \langle u_n \ket{\nabla_{\mathbf{k}}u_n}\cdot d\mathbf{k}
    +\frac{1}{2}\oint_C \frac{\bra{u_n}(\nabla_{\mathbf{k}}H_{\text{AB}})H_{\text{BA}}\ket{u_n}}{1/4-\epsilon_n^2} \cdot d\mathbf{k}
    \Big{]}
    \notag \\
    =&(-1)^m\quad (m\in\mathbb{Z}).
    \label{edge_3DC_revise}
\end{align}

\end{document}